\documentclass[%
 aip,
 amsmath,amssymb,
preprint,%
]{revtex4-1}

\usepackage{graphicx}
\usepackage{dcolumn}
\usepackage{bm}

\usepackage[utf8]{inputenc}
\usepackage[T1]{fontenc}
\usepackage{mathptmx}
\usepackage{etoolbox}
\usepackage{xcolor}
\usepackage[switch]{lineno}
\usepackage{multirow}
\usepackage{ragged2e}

\newcommand{\bb}[1]{\textcolor{black}{#1}}

\newcommand{\mm}[1]{\textcolor{black}{#1}}

\makeatletter
\def\@email#1#2{%
 \endgroup
 \patchcmd{\titleblock@produce}
  {\frontmatter@RRAPformat}
  {\frontmatter@RRAPformat{\produce@RRAP{*#1\href{mailto:#2}{#2}}}\frontmatter@RRAPformat}
  {}{}
}%
\makeatother
\begin{document}

\preprint{AIP/123-QED}

\title[]{Structural transformations in Cu, Ag, and Au metal nanoclusters}
\author{Manoj Settem}
\affiliation{Dipartimento di Ingegneria Meccanica e Aerospaziale, Sapienza Università di Roma, via Eudossiana 18, 00184 Roma, Italy.}

\author{Cesare Roncaglia}
\affiliation{Dipartimento di Fisica dell'Università di Genova, via Dodecaneso 33, 16146 Genova, Italy.}

\author{Riccardo Ferrando$^{\ast}$}
\affiliation{Dipartimento di Fisica dell'Università di Genova and CNR-IMEM, via Dodecaneso 33, 16146 Genova, Italy.}
\email{ferrando@fisica.unige.it; alberto.giacomello@uniroma1.it}

\author{Alberto Giacomello$^{\ast1}$}


\begin{abstract}
Finite-temperature structures of Cu, Ag, and Au metal nanoclusters are calculated in the entire temperature range from 0 K to melting using a computational methodology that we proposed recently [Settem \emph{et al.}, Nanoscale, 2022, 14, 939]. In this method, Harmonic Superposition Approximation (HSA) and Parallel Tempering Molecular Dynamics (PTMD) are combined in a complementary manner. HSA is accurate at low temperatures and fails at higher temperatures. PTMD, on the other hand, effectively samples the high temperature region and melting. This method is used to study the size- and system-dependent competition between various structural motifs of Cu, Ag, and Au nanoclusters in the size range 1 to 2 nm. Results show that there are mainly three types of structural changes in metal nanoclusters depending on whether a solid-solid transformation occurs. In the first type, global minimum is the dominant motif in the entire temperature range. In contrast, when a solid-solid transformation occurs, the global minimum transforms either completely to a different motif or partially resulting in a co-existence of multiple motifs. Finally, nanocluster structures are analyzed to highlight the system-specific differences across the three metals. 
\end{abstract}

\maketitle


\section{Introduction}
Metal nanoclusters constitute an important branch of nanotechnology which exhibit size- and shape-dependent properties. Typically, metal nanoclusters adopt\cite{baletto2005rev} either the non-crystalline icosahedron (Ih) and decahedron (Dh) motifs or the crystalline octahedron (fcc) motif; with the non-crystalline structures being dominant at smaller sizes, but becoming unfavorable at large sizes due to the stress contribution to the energy that is proportional to the volume.\cite{baletto2002potParams,rahm2017coExist,Nelli2023apx} Since properties of technological interest (\bb{catalytic}, optical, etc.) depend on the cluster structure, it is crucial to understand the equilibrium structures of metal nanoclusters. For this purpose, computer simulations can be very useful. Most of the studies available in the literature focus on finding the global energy minimum at a given size.\cite{rahm2017coExist,michaelian1999AuRelSRAmor,grigoryan2005NiCuAu60,shao2005Ag80,apra2006AuRelSRAmor,grigoryan2006Cu150,yang2007Ag160,alamanova2007Ag150,angulo2008Ag,itoh2009CuAg75,bao2009AuBHMC,huang2011Ag141to310,chen2013Ag99,grigoryan2013AgCu150} Although this information is important, it is limited in the sense that global minima refers to the structures at 0 K. However, metal nanoclusters are expected to be produced and observed at finite temperatures. In addition, various structural motifs coexist\cite{schebarchov2018AuHSA,settem2022AuPTMD} at a specific size and temperature. Hence, a method to reliably calculate the equilibrium distribution of various structural motifs in the entire temperature range is essential.

One possible approach is the Harmonic Superposition Approximation (HSA)\cite{hsa1993,calvo2002} which has been used to study Lennard Jones,\cite{doye2001hsaLJ,doye2002hsaLJ,mandelshtam2006hsaLJ,sharapov2007hsaLJ} metal,\cite{schebarchov2018AuHSA,grigoryan2019hsaCu} and alloy nanoclusters.\cite{panizon2015hsaPdPt,bonventre2018hsaAgCuANi} Briefly, in this method, a large number (> 10\textsuperscript{3}) of low-lying minima are sampled from the potential energy surface (PES) to construct an approximation of the partition function. Subsequently, the temperature-dependent probability of an isomer is calculated based on the partition function. HSA captures the structural distribution at low temperatures fairly accurately. However, at higher temperatures, HSA becomes progressively erroneous. This stems mainly from the failure to accommodate the anharmonic effects which become significant at larger temperatures. Another issue is the difficulty in capturing the melting region. In order to reconstruct the melting region, it is necessary to sample the high energy region of the PES which would require one to collect a prohibitively large number of minima. Due to these constraints melting cannot be reliably captured using HSA.

Alternatively, to sample the phase space effectively, one can simulate several \emph{replicas}\cite{earl2005PT} of the system that are at different temperatures and are allowed to exchange configurations at specific intervals according to a Metropolis-like criterion. This method is referred to as \emph{replica exchange} or \emph{parallel tempering}. At higher temperatures, the barriers between various structures are easily overcome ensuring a good sampling at these temperatures. On the other hand, exchange of configurations allows the high temperature configurations to cascade to lower temperatures and, in the process, to improve the phase space exploration at lower temperatures as well. Both Monte Carlo\cite{neirotti2002ptmcLJ1,neirotti2002ptmcLJ2,ballard2014,guimarães2020ptmcSolvation} and molecular dynamics\cite{calvo2012PTMDnFeMelting,tarrat2018PTMDnAu} can be carried out in conjunction with parallel tempering. In PTMC, generally, random displacement moves are employed to sample configurations; which reduces the likelihood of inter-motif transition with increasing cluster size.\cite{nelli2023FarDiss} Also, collective atomic rearrangements\cite{nelli2020AuBHMC} are involved during inter-motif transition involving metallic clusters which might not be straightforward to incorporate into Monte Carlo sampling. As a result, in this work, we carry out parallel tempering with molecular dynamics.

Recently, we have proposed a method\cite{settem2022AuPTMD} that combines HSA and parallel tempering leveraging the advantages offered by these two methods to capture the structural distribution in the entire temperature range (0 K to melting). First, we carry out parallel tempering molecular dynamics (PTMD) with several replicas at temperatures ranging from room temperature to beyond melting. A large collection of local minima are sampled during the PTMD simulations which are then fed into the HSA calculations. This combined method offers several advantages where HSA and PTMD act in a complementary fashion. The conventional HSA calculations require collection of a large number of local minima which are obtained using structure optimization methods.\cite{panizon2015hsaPdPt,schebarchov2018AuHSA} In our case, the minima are directly obtained from PTMD simulations without the need to explicitly search for them. HSA can capture the low temperature solid-solid transitions which might prove to be elusive for PTMD. On the other hand, PTMD captures the high temperature and the melting regions accurately where HSA calculations fail. As a result, the low temperature and the high temperature regions are accurately captured by HSA and PTMD respectively. In the intermediate temperatures, HSA and PTMD have a good agreement.

In this work, we apply this method to study the size- and system-dependent structural changes with temperature in Cu, Ag, and Au metal nanoclusters. This is crucial information given their strong influence on the properties of metal nanoclusters. For example, catalytic activity of metal nanoclusters depends on the structure type and size\cite{li2015AuCatShapeEffect,zhao2017sizeCatCu,jørgensen2018siteAssembly,rong2021sizeDepCat} due to the wide variety of catalytic sites.\cite{rossi2019npGenome} In addition, the catalytic activity can be enhanced by an ensemble of different geometrical structures in comparison to homogeneously shaped structures.\cite{cheula2020npEnsembleCat} Hence, it is essential to gather knowledge on the equilibrium structural distribution where various geometrical motifs coexist.

Several theoretical works have calculated the global minimum structures of Cu, Ag, and Au nanoclusters. Grigoryan \emph{et al.}\cite{grigoryan2006Cu150} calculated the global minima of Cu clusters up to 150 atoms using the embedded atom method (EAM),\cite{DawPRB1984} and up to 60 atoms using Gupta\cite{gupta1981} and Sutton-Chen\cite{suttonChen1990} potentials. Highly stable structures occur at the sizes 13, 19, 55, 92, and 147 with all of them having high symmetry icosahedral structures except 92 which is a chiral structure having \emph{T} point group symmetry. Most of the structures are icosahedra with the sizes 4, 17, 26, 28, 29, 91$-$95 having tetrahedral geometry and 75, 78, 81, 101$-$103 being decahedra. In the case of Ag nanoclusters of sizes larger than 60 atoms, decahedron is found to be the dominant motif.\cite{shao2005Ag80,yang2007Ag160,alamanova2007Ag150,huang2011Ag141to310} There are few exceptions where truncated octahedron (fcc) and icosahedron (Ih) are the global minima. Due to the strong relativistic effects,\cite{pyykko2004AuRelEffects} Au nanoclusters exhibit peculiar structures. At sizes smaller than 40 atoms, Au nanoclusters adopt either planar or hollow cage-like geometries.\cite{furche2002AuPlanarStructsExpIon,hakkinen2002Au2Dto3DCuAg,hakkinen2003AuPlanarStructs,johansson2004AuCageStructs,gu2004AuCageStructs,fa2006AuCageStructs,xing2006AuCageStructs} In comparison to Cu and Ag, Au disfavors icosahedral structures. At the magic sizes of 55, 147, and 309 the icosahedron is not the global minimum.\cite{bao2009AuBHMC,schebarchov2018AuHSA,nelli2020AuBHMC} This is also evident over larger size range (up to 1000 atoms).\cite{rahm2017coExist} However, when the icosahedral structures are observed in Au nanoclusters, for example, at higher temperatures,\cite{mottet2005rosette,rossi2018structTrans} they typically have ``rosette''\cite{apra2004AuRosette,Nelli2022epjap} defects on the surface. A ``rosette'' defect appears when a vertex atom is pushed out to form a six-atom ring with the five neighboring surface atoms leaving behind a vacancy at the vertex position.

Cu, Ag, and Au clusters have also been studied using density functional theory (DFT) calculations. Generally, ideal structures are considered since global minimum search becomes prohibitive at the DFT level for clusters larger than $\sim$ 50 atoms.\cite{yin2021coinageRev} Roldán \emph{et al.}\cite{roldan2008DFTfcc} carried out structural analysis of several ``magic'' sized octahedral Cu, Ag, and Au clusters in the range 38 $-$ 225 atoms and identified a correlation to estimate cohesive energies in a large size range. Similarly, Kiss \emph{et al.}\cite{kiss2011AgDFTSizeEffects} studied octahedral and icosahedral Ag clusters (consisting of 6 $-$ 600 atoms) and observed that the cohesive energy is linear with inverse of cluster size. Oliveira \emph{et al.}\cite{oliveira2016benchmarkingDFTB} showed that Ag icosahedra are energetically stable compared to cuboctahedra through density functional tight binding (DFTB) calculations of ``magic'' clusters in the range 55 $-$ 561 atoms.

The picture arising from experiments is more complex, since in experiments it is often difficult to disentangle kinetic effects from equilibrium ones.\cite{baletto2005rev} Electron microscopy has been used to study the structure of metal nanoclusters with varying size and temperature. Langlois \emph{et al.}\cite{langlois2008CuHRTEM} prepared Cu nanoparticles in a broad size range of 1 nm to 12 nm using thermal evaporation. They observed a significant overlap between icosahedra and decahedra at sizes less than 8 nm beyond which \emph{fcc} structures were observed. Volk \emph{et al.}\cite{volk2013AgTEM} analyzed Ag clusters with size < 7 nm grown in superfluid He droplets. The smallest particles were fcc, with decahedra at intermediate sizes and icosahedra at large sizes. However, theoretical predictions\cite{baletto2002potParams,rahm2017coExist} show that icosahedra are energetically favored at smaller sizes while fcc are favored at larger sizes, while large icosahedra are likely to be due to kinetically trapped growth on top of smaller decahedra.\cite{Baletto2001prb,Elkoraychy2022nh} 

Recently, the structural distribution of size-selected Ag clusters centered around 309 atoms was measured,\cite{loffreda2021Ag309TEM} finding an abundance of fcc structures with very little icosahedra (2\%). This is in contrast to the prediction that icosahedra is the dominant motif around the size 309.\cite{rahm2017coExist} Wells \emph{et al.}\cite{wells2015AuImagingACFraction} calculated the proportion of various motifs of Au$_{561}$, Au$_{742}$, and Au$_{923}$. At these sizes, fcc and decahedra making up 70\% of the structures while icosahedra contribute less than 5\%. Finite-temperature distribution of Au$_{561}$ was calculated by Foster \emph{et al.}\cite{foster2018AuImagingACFraction} in the temperature range 20 $^\circ$C to 500 $^\circ$C. Again, icosahedra were almost non-existent beyond 100 $^\circ$C with less than 3\%. At temperatures greater than 125 $^\circ$C, there is an increase in the proportion of decahedra at the expense of fcc structures. The experiments establish a lack of preference for the icosahedral motif in Au nanoclusters, in agreement with the findings of Gupta potential and DFT calculations.\cite{palomaresBaez2017AuDFT}


From a theoretical and experimental viewpoint, it is essential to have a knowledge of the equilibrium proportion of various structural motifs as a function of temperature. In this work we calculate the structural distribution of Cu, Ag, and Au metal nanoclusters at the sizes 90, 147, and 201 which fall in the size range of 1 nm to 2 nm. These were chosen to highlight the size- and system-dependent structural changes. \bb{147 and 201 are ``magic'' sizes corresponding to perfect icosahedron (147) and regular truncated octahedron (201). It is generally assumed that ``magic'' sized structures have energetic stability. Our results show that this assumption is not always true. Finally, we chose 90 to look at non-magic sized structures.}

\section{Methods}
We use the tight binding model within the second moment approximation (TBSMA)\cite{tbsma1971} which is also referred to as Gupta\cite{gupta1981} potential or Rosato-Guillope-Legrand (RGL)\cite{rgl1989} potential to model the atom-atom interactions in Cu, Ag, and Au nanoclusters. The parameters of the Gupta potential have been taken from Ref.\cite{baletto2002potParams}. The interaction potential of Au gives an accurate description of the experimental cluster structures in gas phase\cite{wells2015AuImagingACFraction} and on MgO substrates.\cite{han2014imagingOnMgO} In addition, this potential agrees well with DFT calculations in the prediction of surface ``rosette'' defects in icosahedra\cite{apra2004AuRosette} and the tendency to disfavor icosahedra.\cite{palomaresBaez2017AuDFT} Coming to Ag and Cu, the Gupta potentials correctly predict the stability of Mackay stacking over anti-Mackay stacking in icosahedral clusters in line with the DFT calculations (see Supporting Information in ref. \cite{BochicchioChiral2010}). In Ag$_{586}$, fcc structure is energetically preferred in comparison to icosahedron which is also the case according to DFT.\cite{panizon2014diluteImpurity}

Gupta potential predicts correctly that Ag icosahedra are energetically stable compared to cuboctahedra which agrees with the DFTB calculations\cite{oliveira2016benchmarkingDFTB} (see the plot of energy difference between cuboctahedron and icosahedron in supplementary figure S1). At the size 147, icosahedra and decahedra are the prominent motifs. In order to assess the competition between these motifs, we have carried out DFT calculations for Cu$_{147}$ and Ag$_{147}$. For Au$_{147}$ clusters, we refer to the calculations done previously.\cite{palomaresBaez2017AuDFT} DFT calculations were carried out using Quantum ESPRESSO\cite{quantumEspresso} code. Projected augmented wave (PAW)\cite{kresse1996VASPcms} pseudopotentials were used with Perdew-Burke-Ernzerhof (PBE)\cite{pbeXCFunctional} exchange-correlation functional. An energy cutoff of 45 Ry was used for both Ag, Cu; while the charge density cutoff of 181 Ry, 236 Ry were used for Ag, Cu respectively. The calculations were considered to be converged with energy and force tolerance of $1
\times10^{-4}$ Ry and $1
\times10^{-3}$ Ry/a.u. respectively. The energy difference between decahedron (Dh) and icosahedron (Ih) defined as, $E_{Dh}-E_{Ih}$, at the DFT/PBE level are +3.87 eV, +2.55 eV, and -2.56 eV for Cu, Ag, and Au respectively. The corresponding values according to Gupta potential are +1.57 eV, +0.46 eV, and -1.86 eV. Both DFT/PBE and Gupta show therefore the same trend: Ih is energetically preferred in Cu and Ag while Dh is favored in Au.
Based on these results, we believe that Gupta potentials are reliable for analyzing structural trends between Cu, Ag, and Au metal nanoclusters. The use of this model will allow a thorough sampling of the energy landscape which would be hardly feasible by DFT. \mm{A detailed comparison of Gupta potential with DFT calculations is provided in the \emph{Results and Discussion} section which allows us to assess its performance and limitations.}

Before the PTMD simulations, we calculate the global minimum at each size using basin hopping Monte Carlo (BHMC)\cite{nelli2020AuBHMC,rossi2009BHMC,settem2022AuPTMD} optimization search. For each size, we run five independent search simulations with at least 2.5$\times$10\textsuperscript{5} basin hopping steps.

The detailed procedure of the combined method of PTMD+HSA is described in a previous work\cite{settem2022AuPTMD}. Here we only recapitulate it briefly. In the PTMD simulations, there are two fundamental parameters: the number of replicas ($M$) and the temperature, T\textsubscript{m} ($m=1,2,3,...,M$) of each replica. All the replicas are in a canonical ensemble (\emph{NVT}) and exchange of configurations between a pair of replicas is attempted at specific intervals. The number of replicas is chosen such that we have at least 20$-$30\%  acceptance of the replica swaps. This is achieved by calculating an approximate caloric curve to identify the melting range and then adjusting the number of replicas and their temperatures to achieve the desired swap acceptance rate. All the PTMD simulations have been carried out in LAMMPS.\cite{lammps} We use a time step of 5 fs for the molecular dynamics evolution and replica swaps are attempted every 250 ps. They are either accepted or rejected according to a Metropolis-like criterion.

We begin the PTMD simulations with all the replicas having the same structure, either global minimum or a low energy structure. After discarding the initial phase of PTMD ($\sim$ 0.5 $\mu$s), we sample configuration at 125 ps after a swap attempt for a total time of about 1 $\mu$s to 2 $\mu$s. 

The configurations sampled from PTMD simulations are also fed into the HSA analysis. In the HSA\cite{schebarchov2018AuHSA,panizon2015hsaPdPt,bonventre2018hsaAgCuANi} method, the partition function is given by,
\begin{equation} \label{eqnPartFunc}
    Z=\sum_{i}\frac{e^{-\beta{E^0_i}}Z^{tr}_iZ^{rot}_iZ^{vib}_i}{g_i}
\end{equation}
where $\beta=1/(k_BT)$. The summation is over all the local minima, $i$, considered for the HSA. $E^0_i$ is the energy of the local minimum, $i$. $Z^{tr}$, $Z^{rot}$,  and $Z^{vib}$ are the translational, rotational, and vibrational contributions to the partition function, respectively. It has been shown that only the vibrational contribution is sufficient to calculate the probability of the local minima.\cite{panizon2015hsaPdPt} The denominator, $g_i$, is the order of the symmetry group of the local minimum $i$. The vibrational contribution due to a single minimum is given by
\begin{equation} \label{eqnPartFuncVib}
    Z^{vib}=\prod_{n=1}^{3N-6}\frac{e^{-{\beta}{\hbar}{\omega}_n/2}}{1-e^{-{\beta}{\hbar}{\omega}_n/2}} 
\end{equation}
where $\omega_n$ are the $3N-6$ ($N$ is the number of atoms in the cluster) frequencies of the normal modes. The probability of a local minimum as a function of temperature is now given by
\begin{equation} \label{eqnProbHSA}
    p_i=\frac{e^{-\beta{E^0_i}}Z^{vib}_i/g_i}{\sum_{j}e^{-\beta{E^0_j}}Z^{vib}_j/g_j} 
\end{equation}
We define the probability of a specific structure type ($p^{struct}$) by summing up the probabilities of all the minima belonging to that structure type.
\begin{equation} \label{eqnProbHSAStruct}
    p^{struct}=\sum_{k}p_k
\end{equation}
where $k$ represents all the minima having the same structure. Local minima for the HSA analysis were collected from PTMD simulations up to an energy cutoff of 1 eV to 1.5 eV \mm{with the exception of Cu$_{147}$ and Ag$_{147}$ where 2.5 eV was used}. Two minima were considered to be different if they belonged to different structure types and were separated by at least 0.05 meV in energy. For identifying the geometrical motif of a given configuration, we use common neighbor analysis (CNA)\cite{cna1994} signatures. The structures are classified using the same scheme that we employed for Au nanoclusters previously\cite{settem2022AuPTMD,Roncaglia2021pccp} and categorize them into decahedron (Dh), icosahedron (Ih), twin, fcc, and amorphous structure classes. A structure that does not fall into any of these categories is classified as a \emph{mix} structure. Typically, these structures are not well defined or contain structural features of more than one geometrical motif. These structures will be described in more detail while presenting the results. \mm{Further details about the parameters used for HSA and PTMD are provided in the Supplementary Information.}

\section{Results and Discussion}
We will present the results of Cu and Ag nanoclusters. We note that structural distribution of Au nanoclusters has been previously reported by us\cite{settem2022AuPTMD} and we use it here to make a comparison with Cu and Ag. Also, we compare in detail the structures of Au, Cu, and Ag which was not reported previously. To begin with, we discuss the finite-temperature structural distributions and then make a comparison to highlight the differences and similarities between Cu, Ag, and Au clusters. \bb{The melting point of all the metal nanoclusters in the current work are reported in Table \ref{tab:mp_table}. We identify the melting point by first constructing the heat capacity (C$_V$) curve from PTMD simulations. Melting point is then calculated as the peak of C$_V$ curve.}

\begin{table}[!h]
\small
\caption{\bb{Melting point (in K) of Cu, Ag, and Au nanoclusters.}}
\centering
{\def\arraystretch{1.15}
\begin{tabular*}{0.95\textwidth}{@{\extracolsep{\fill}}ccc|ccc|ccc}
\hline
{Cu$_{90}$} & {Cu$_{147}$} & {Cu$_{201}$} & {Ag$_{90}$} & {Ag$_{147}$} & {Ag$_{201}$} & {Au$_{90}$} & {Au$_{147}$} & {Au$_{201}$} \\
\hline
{609} & {779} & {745} & {510} & {651} & {654} & {420} & {505} & {550}\\
\hline
\end{tabular*}
}
\label{tab:mp_table}
\end{table}

\subsection{Cu}
\begin{figure}[!t]
\centering
  \includegraphics[width=1.0\textwidth]{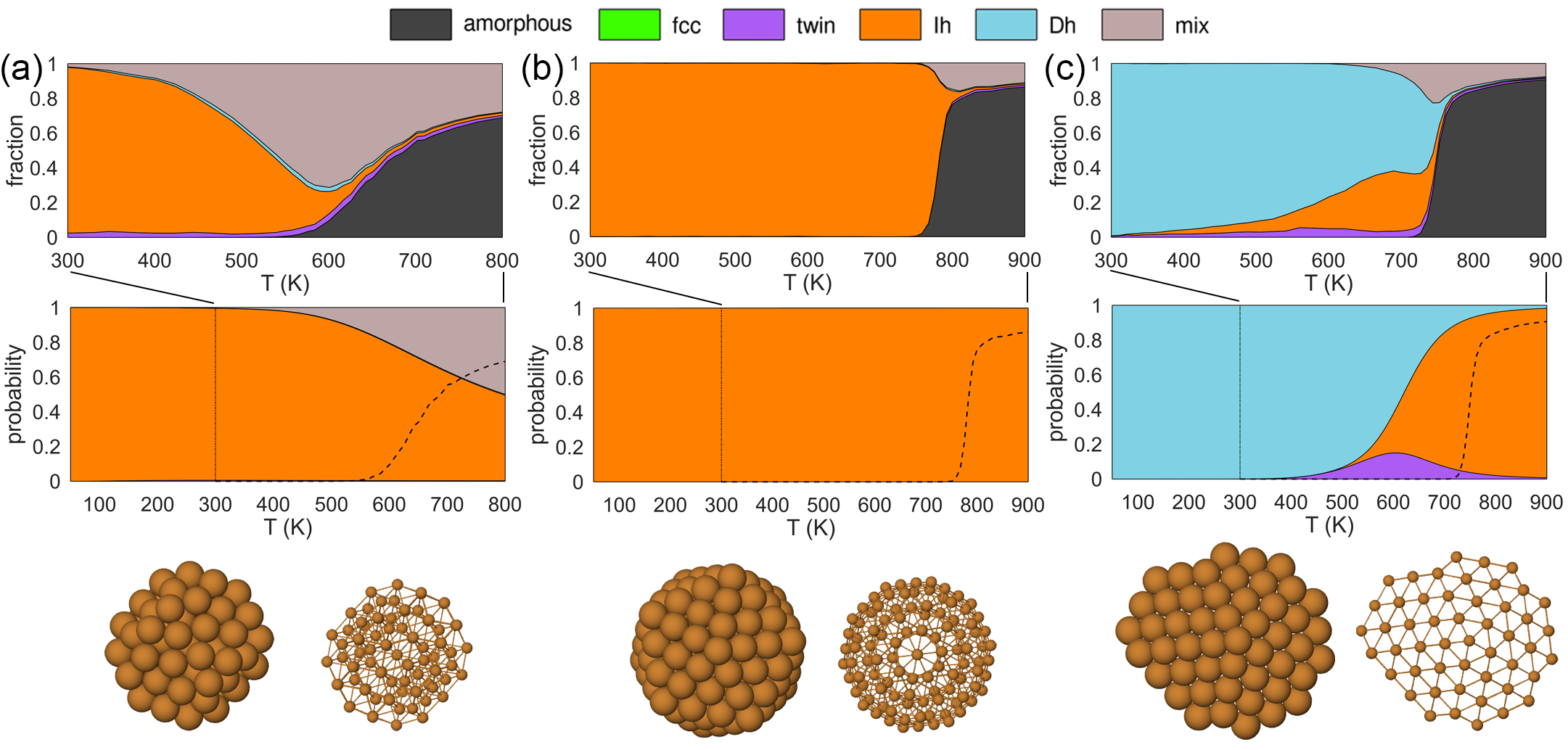}
  \caption{Structural distribution of (a) Cu$_{90}$, (b) Cu$_{147}$, and (c) Cu$_{201}$. PTMD, HSA results are shown in the top and middle rows. Global minimum structures are shown in the bottom row. In the HSA results, for comparison, we report with vertical lines  the range of PTMD temperatures and with a dashed line the fraction of amorphous structures calculated from PTMD simulations.} 
  \label{fgr:cu_ptmd_hsa}
\end{figure}

Cu has a strong preference for icosahedral motif as compared to Ag and Au.\cite{baletto2002potParams,rahm2017coExist} The global minimum of Cu$_{90}$, Cu$_{147}$, and Cu$_{201}$ are shown in Fig. \ref{fgr:cu_ptmd_hsa}. The global minimum of Cu$_{90}$ and Cu$_{147}$ are both icosahedra with Cu$_{90}$ having \emph{C\textsubscript{2v}} point group symmetry. However, with the EAM potential, the global minimum of Cu$_{90}$ was predicted to be an icosahedron with \emph{C\textsubscript{s}} symmetry.\cite{grigoryan2006Cu150} The best structure of Cu$_{201}$ is a decahedron with \emph{C\textsubscript{s}} point group symmetry.

In the case of Cu$_{90}$, icosahedron (Ih) is the dominant motif at room temperature with very small amount of twins, decahedra (Dh) and \emph{mix} structures (Fig. \ref{fgr:cu_ptmd_hsa}a). The \emph{mix} structures comprise several different geometric types. Predominantly, the \emph{mix} structures consist of icosahedral-based geometries that either have amorphous regions or the entire structure adopts a configuration similar to the 92-atom chiral structure\cite{grigoryan2006Cu150,settem2020chiralCuCore} with two missing atoms. The remaining \emph{mix} structures consist of polydecahedra (p-Dh) which have more than one local fivefold axis.\cite{polyDh2007,nelli2020AuBHMC,settem2022AuPTMD} With increasing temperature, the proportion of \emph{mix} structures increases at the expense of Ih and peaks before melting at $\sim$ 600 K. Qualitatively, HSA predicts similar structural changes in Cu$_{90}$. The agreement between HSA and PTMD is good at room temperature and thereafter there are quantitative discrepancies. The increase in \emph{mix} structures is rather slow according to HSA. For example, at 600 K, PTMD predicts 71.2\% \emph{mix} structures, while HSA predicts only 20.6\%.

At size 147 (Fig. \ref{fgr:cu_ptmd_hsa}b), the icosahedron, which is the global minimum, dominates in the entire temperature range according to both PTMD and HSA. This indicates a high thermal stability of the icosahedral motif at this size. Moving on to Cu$_{201}$ (Fig. \ref{fgr:cu_ptmd_hsa}c), again, the global minimum structure, a decahedron, dominates at room temperature and its proportion decreases steadily with temperature. Icosahedra compete with decahedra at higher temperatures with maximum proportion of Ih observed at 700 K just before melting. HSA on the other hand, predicts a significantly higher amount of Ih at this temperature (77.2\% vs 33.7\%). Interestingly, fcc and twin structures are almost absent in Cu$_{90}$, Cu$_{147}$, and Cu$_{201}$ clusters in the entire temperature range.

\subsection{Ag}
\begin{figure}[!b]
\centering
  \includegraphics[width=1.0\textwidth]{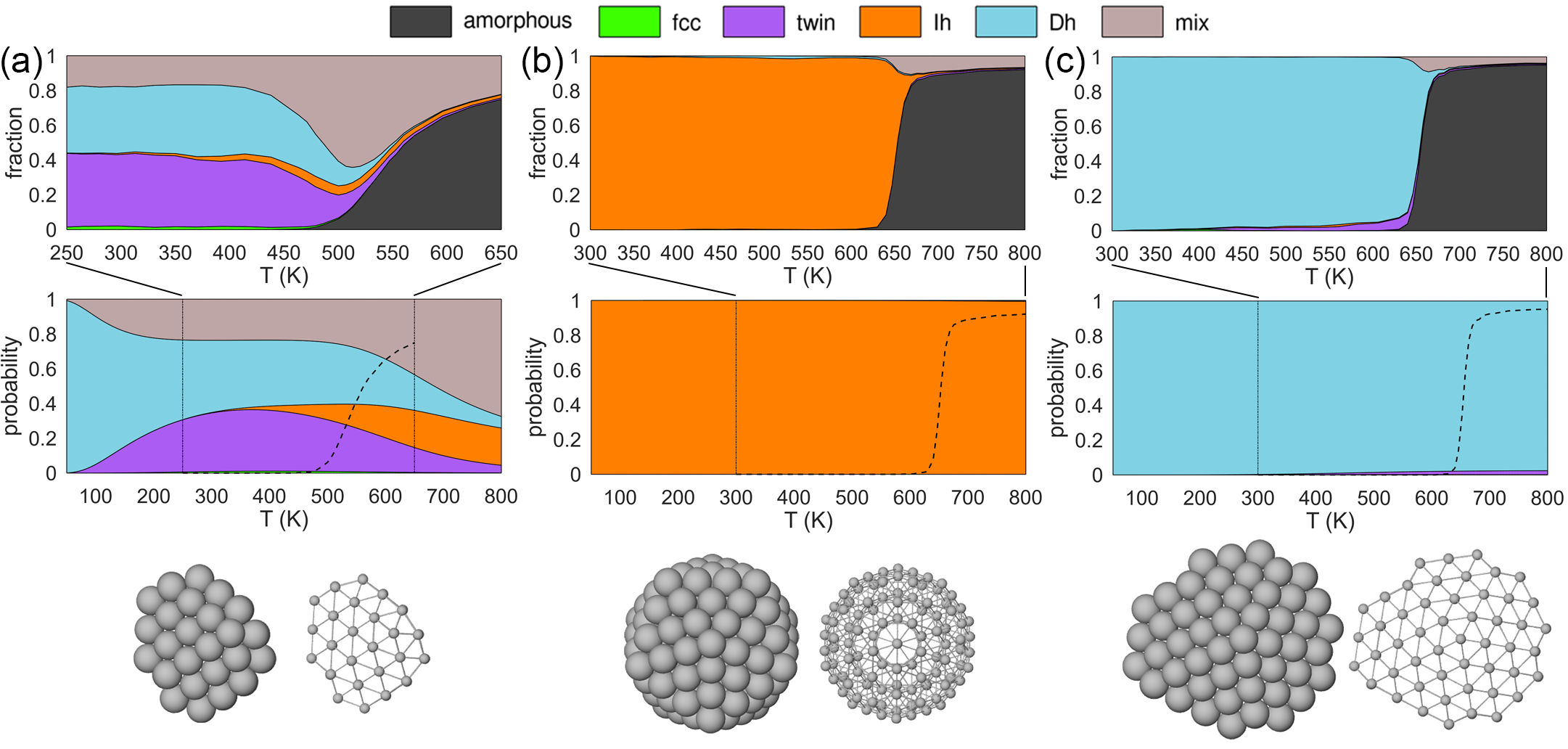}
  \caption{Structural distribution of (a) Ag$_{90}$, (b) Ag$_{147}$, and (c) Ag$_{201}$. PTMD, HSA results are shown in the top and middle rows. Global minimum structures are shown in the bottom row. In the HSA results, for comparison, we report with vertical lines  the range of PTMD temperatures and with a dashed line the fraction of amorphous structures calculated from PTMD simulations.} 
  \label{fgr:ag_ptmd_hsa}
\end{figure}

The global minimum structures of Ag$_{90}$ and Ag$_{201}$ (Fig. \ref{fgr:ag_ptmd_hsa}) are decahedra with both structures having \emph{C\textsubscript{s}} point group symmetry. The ideal icosahedron is the global minimum of Ag$_{147}$. These results are consistent with the previously reported global minima at these sizes for Ag clusters.\cite{yang2007Ag160,alamanova2007Ag150,huang2011Ag141to310}

Ag$_{90}$ exhibits interesting structural changes (Fig. \ref{fgr:ag_ptmd_hsa}a). From the HSA results, it is evident that the global minimum decahedron undergoes a partial transition to twin and \emph{mix} structures with increasing temperature, leading to a combination of Dh $+$ twin $+$ \emph{mix} structures at 250 K. Considering the PTMD results, the proportion of Dh, twins, and \emph{mix} structures remains constant up to $\sim$ 450 K. This is a case of one-to-many solid-solid transition\cite{panizon2015hsaPdPt} where one geometrical motif, the Dh, is replaced by a coexistence of Dh, twins, and \emph{mix} structures. Above 450 K, the proportion of \emph{mix} structures increases at the expense of Dh and twins. The \emph{mix} structures are a combination of polydecahedra\cite{nelli2020AuBHMC} and distorted icosahedra having amorphous regions. The structural changes in Ag$_{147}$ and Ag$_{201}$ (Figs. \ref{fgr:ag_ptmd_hsa}b, c) are fairly straightforward. In both cases, the global minimum motif (Ih for 147 and Dh for 201) dominates in the entire temperature range, with other motifs nonexistent or in extremely small proportions.

\subsection{Au}
We have recently\cite{settem2022AuPTMD} reported the structural changes in Au nanoclusters and hence, we will only summarize them briefly here (see supplementary figure S2). The global minimum structures of Au$_{90}$, Au$_{147}$, and Au$_{201}$ are fcc, decahedron, and fcc (ideal truncated octahedron), respectively. At size 90, the global minimum motif, fcc, is dominant at lower temperatures and competes with twin and \emph{mix} structures. With increasing temperature, fcc structures decrease along with an increase in \emph{mix} structures. In the case of Au$_{147}$, the decahedron (global minimum) remains dominant up to melting along with small amounts of twin and fcc structures. Above 400K, Ih and \emph{mix} structures begin to appear with \emph{mix} structures dominating close to melting. In Au$_{201}$, there is a solid-solid transition from the fcc global minimum (a truncated octahedron) to a Dh at low temperature around 200 K. Thereafter, the Dh dominates up to melting along with a small amount of twins ($\sim$ 10\%).

\begin{figure}[!t]
\centering
  \includegraphics[width=1.0\textwidth]{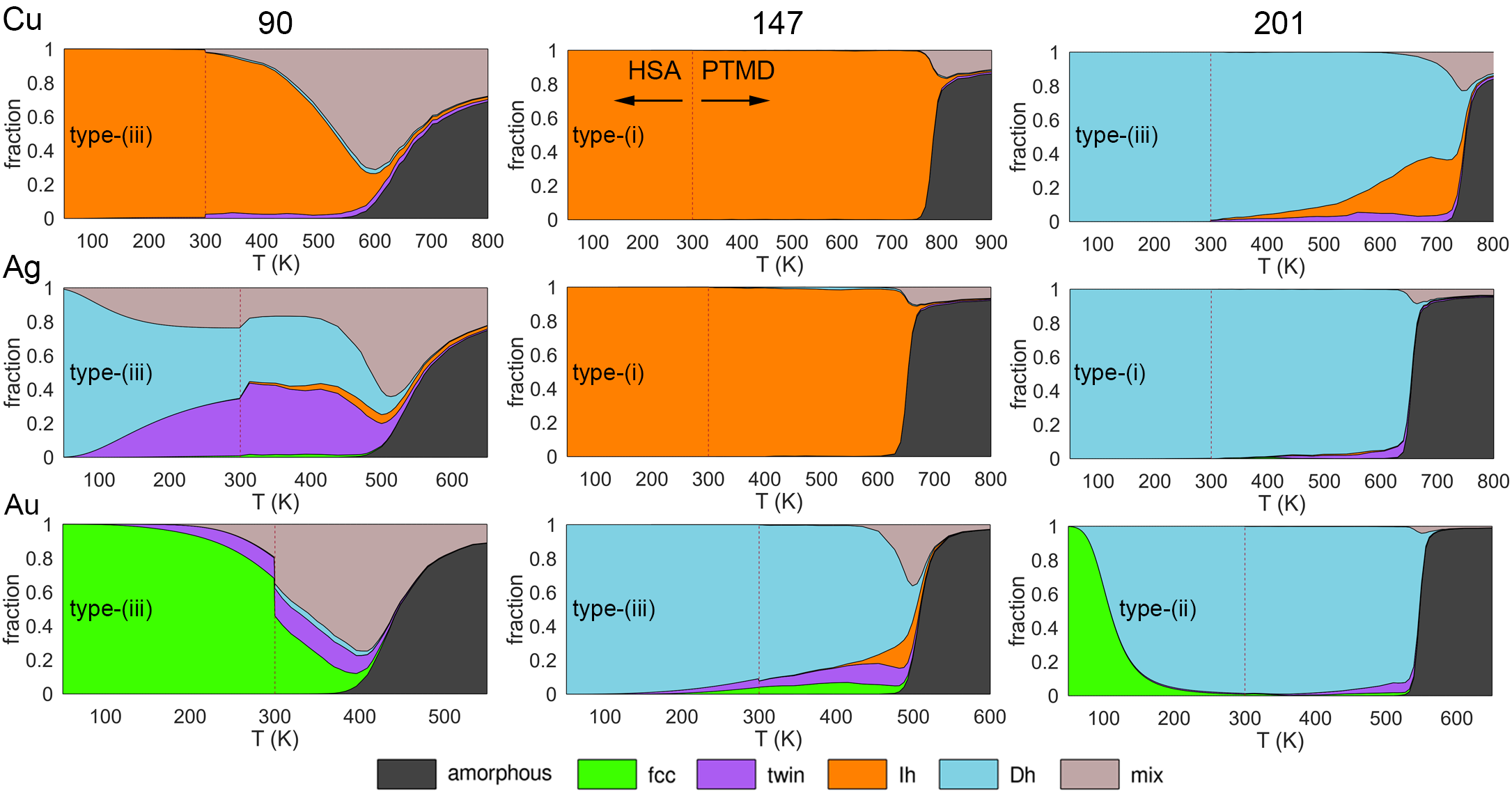}
  \caption{Structural changes in the entire temperature range by combining HSA and PTMD in Cu, Ag, and Au nanoclusters. Vertical line in each plot indicates the temperature at which HSA and PTMD are stitched together. \bb{The type of structural transformation is also indicated. Description of the various types of structural transformations is provided in the text.}}
  \label{fgr:stitch_all}
\end{figure}

\subsection{Cu, Ag, and Au all together: combined HSA+PTMD}
We stitch together HSA and PTMD results in order to get the structural changes in the entire temperature range in a single plot. Data for Au are taken from ref. \cite{settem2022AuPTMD}. Figure~\ref{fgr:stitch_all} compares the available results for Cu, Ag, and Au at all temperature and sizes. HSA and PTMD are stitched together at $300$~K. At the temperature where HSA and PTMD are joined, their structural distributions have an excellent agreement except for small jumps in the case of Ag$_{90}$ and Au$_{90}$, where the trends are anyway consistent. This shows that our approach of combining HSA and PTMD is fairly robust and validated across various metal systems.

There are broadly three categories of structural changes that can be observed: type-(i) the global minimum remains the dominant motif up to melting, where amorphous takes over; type-(ii) solid-solid transitions occur, either completely or partially, well below melting temperature, resulting in an entirely different dominant motif; type-(iii) solid-solid transitions gradually occur leading to a co-existence of multiple motifs. 
The cases Cu$_{147}$, Ag$_{147}$ and Ag$_{201}$ of fall into the first category, while Au$_{201}$ falls into the second category. All other cases fall into the third category, but with some differences. In Au$_{147}$  and Cu$_{201}$, the coexistence between motifs is present in a relatively narrow temperature range close to melting, whereas in all clusters of size 90 coexistence is already found at low temperatures. 

The results show that ideal geometries corresponding to the ``magic'' sizes are not necessarily energetically preferred. Here we considered two ``magic'' sizes, 147 and 201. At size 201, truncated octahedron has the perfect geometry. However, only Au has truncated octahedron as the global minimum, while decahedron prevails for Cu and Ag. Even in Au, the global minimum transforms to Dh which remains the dominant structure at finite temperatures. 
On the other hand, at size 147, which corresponds to a perfect icosahedron, both Cu and Ag have this structure as the global minimum. However, decahedron is the global minimum of Au$_{147}$ with some icosahedra appearing only above 400 K. At size 90, all three systems have a different geometrical motif as the global minimum -- Ih for Cu$_{90}$, Dh for Ag$_{90}$, and fcc for Au$_{90}$ -- which remains dominant (Cu, Au) or competes with other motifs (Ag). The structural distribution of Cu reinforces the strong preference of icosahedral motif in Cu clusters.

\subsection{Structural characterization}
We have, so far, discussed how the various geometrical motifs compete with temperature. In this section, we characterize the structural features of the various motifs.

\begin{figure}[!t]
\centering
  \includegraphics[width=0.7\textwidth]{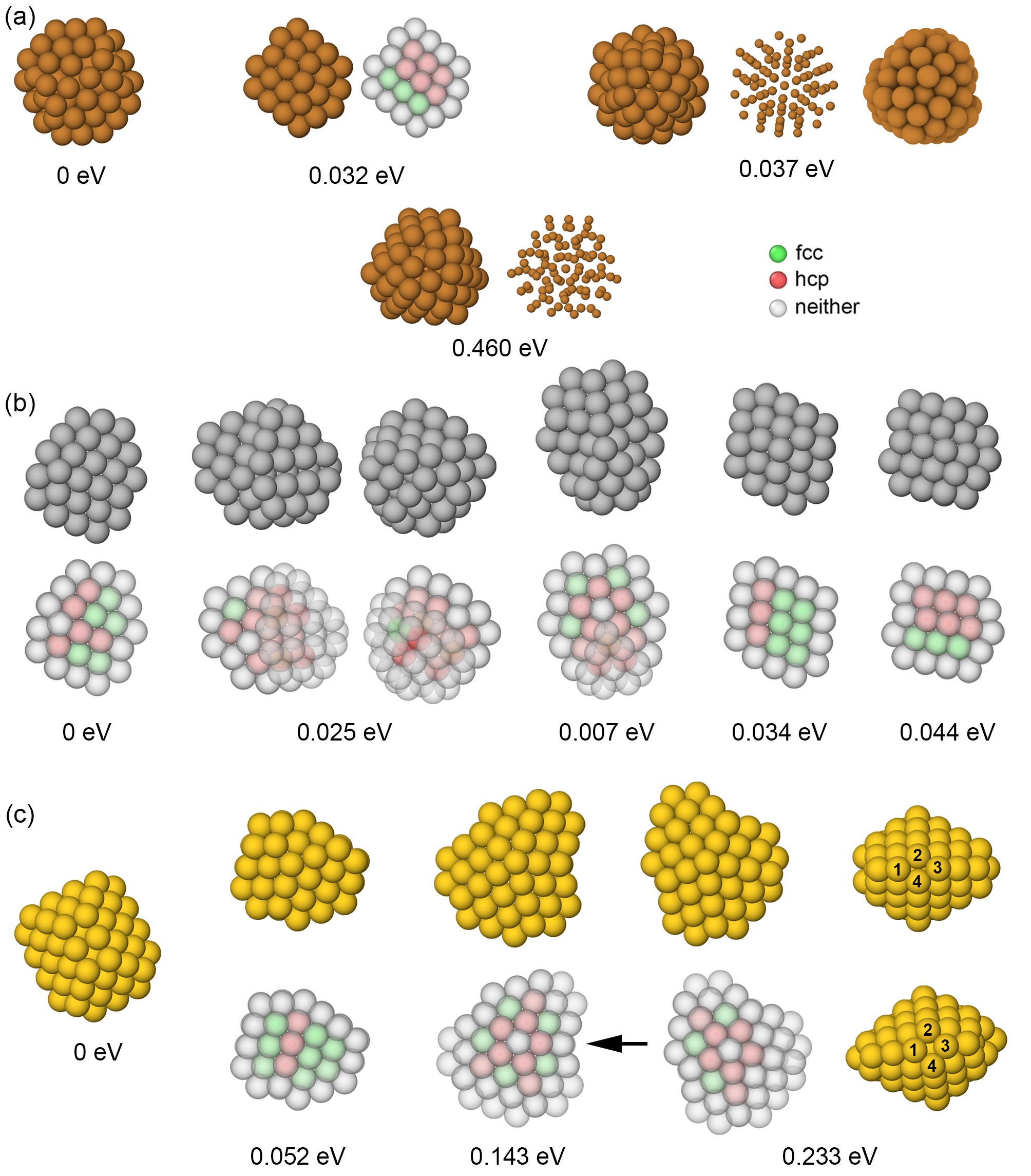}
  \caption{Structures of (a) Cu$_{90}$, (b) Ag$_{90}$, and (c) Au$_{90}$. The energy of each structure is relative to the global minimum (0 eV). The arrow in (c) shows the relatively deeper reentrant groove in Au compared to Cu and Ag. Atoms marked 1, 2, 3, and 4 show the surface restructuring in  Au$_{90}$ decahedron.} 
  \label{fgr:structures_90}
\end{figure}

Typical structures of  Cu$_{90}$ are shown in Fig. \ref{fgr:structures_90} along with their energies relative to the global minimum. The icosahedron is the dominant motif of Cu$_{90}$ along with minor amounts of twin and Dh. The twin structures of Cu$_{90}$ typically have stacking faults (second structure in Fig. \ref{fgr:structures_90}a). At higher temperatures, icosahedra resembling the 92-atom incomplete Mackay icosahedron having \emph{C\textsubscript{3v}} point group symmetry are observed. These structures have two surface vacancies at various positions on the 92-atom cluster resulting in Cu$_{90}$ icosahedra. An example is shown in the  third structure in Fig. \ref{fgr:structures_90}a. As the temperature increases further, some of these  icosahedra undergo a twist and transform to \emph{mix} structures resembling the 92-atom chiral geometry with tetrahedral \emph{T} symmetry (fourth structure in Fig. \ref{fgr:structures_90}a). The 92-atom chiral structure is the global minimum\cite{grigoryan2006Cu150,settem2020chiralCuCore} of Cu$_{92}$ and has also been experimentally confirmed to have \emph{T} symmetry from a comparison of the photoelectron spectra of Na and Cu clusters.\cite{pesNaCu2005,olegThesis,oleg92} Again, the chiral-like Cu$_{90}$ clusters have two surface vacancies.

In the case of Ag$_{90}$ (Fig. \ref{fgr:structures_90}b), along with the conventional decahedron (first structure), we find decahedra with with either one (third structure) or two (second structure) hcp islands. When the two hcp islands are adjacent to each other, a local decahedral axis is formed at the intersection which can be considered as a polydecahedron (p-Dh)\cite{polyDh2007} which has more than one decahedral axis. The twin motif which competes with Dh consists of either a single hcp plane (fourth structure) or stacking faults (fifth structure). Moving on to Au$_{90}$ (Fig. \ref{fgr:structures_90}c), the twins predominantly have a single hcp plane unlike Cu$_{90}$ and Ag$_{90}$. Also, Au$_{90}$ decahedra have deeper reentrant grooves (see the arrow in Fig. \ref{fgr:structures_90}c) compared to decahedra of Cu and Ag. This is consistent with the general trend found in Ref.\cite{baletto2002potParams} The decahedron can undergo surface restructuring resulting in a \emph{mix} structure (see fourth structure in Fig. \ref{fgr:structures_90}c). Consider the four atoms (1, 2, 3, and 4) shown before (top) and after (bottom) restructuring. The atoms 2, 4 are pushed apart and the atoms 1, 3 come closer leading to a \{100\} like arrangement.

\begin{figure}[!t]
\centering
  \includegraphics[width=0.3\textwidth]{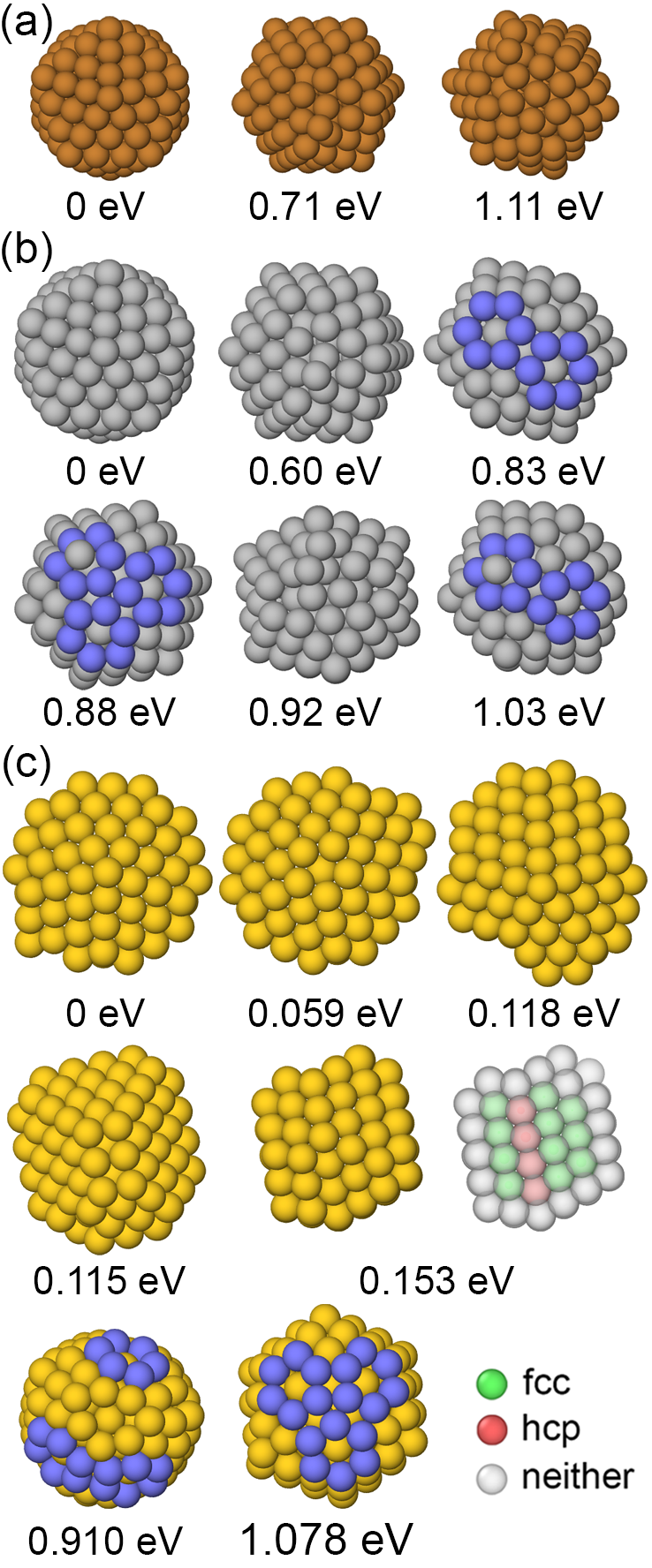}
  \caption{Structures of (a) Cu$_{147}$, (b) Ag$_{147}$, and (c) Au$_{147}$. The energy of each structure is relative to the global minimum (0 eV). ''Rosette'' defects in (b) and (c) are highlighted in blue.} 
  \label{fgr:structures_147}
\end{figure}

At size 147, we observe a gradual change in the nature of icosahedra from Cu to Ag to Au. With increasing temperature, the perfect icosahedra becomes defective with, initially, single vertex vacancy (second structure in Fig. \ref{fgr:structures_147}a) and at still higher temperatures, multiple vacancies (third structure). These same vertex vacancies are also observed in Ag$_{147}$ icosahedra (second and fifth  structures in Fig. \ref{fgr:structures_147}b). However, along with the vertex vacancies, we also observe ``rosette''\cite{apra2004AuRosette,Nelli2022epjap} defects where the vertex atom protrude to join the five nearest neighbors on the surface to form a six-atom ring. These are highlighted in blue for Ag$_{147}$ in Fig. \ref{fgr:structures_147}b where either two or three ``rosette'' defects occur together. Icosahedra in Au$_{147}$, which appear mainly above 400 K, almost always have ``rosette'' defects as shown in Fig. \ref{fgr:structures_147}c. Au$_{147}$ decahedra at higher temperatures exhibit deep reentrant grooves compared to the global minimum (second and third structures in \ref{fgr:structures_147}c). The twins in Au$_{147}$ predominantly have single hcp planes as shown in \ref{fgr:structures_147}c.

\begin{figure}[!t]
\centering
  \includegraphics[width=0.3\columnwidth]{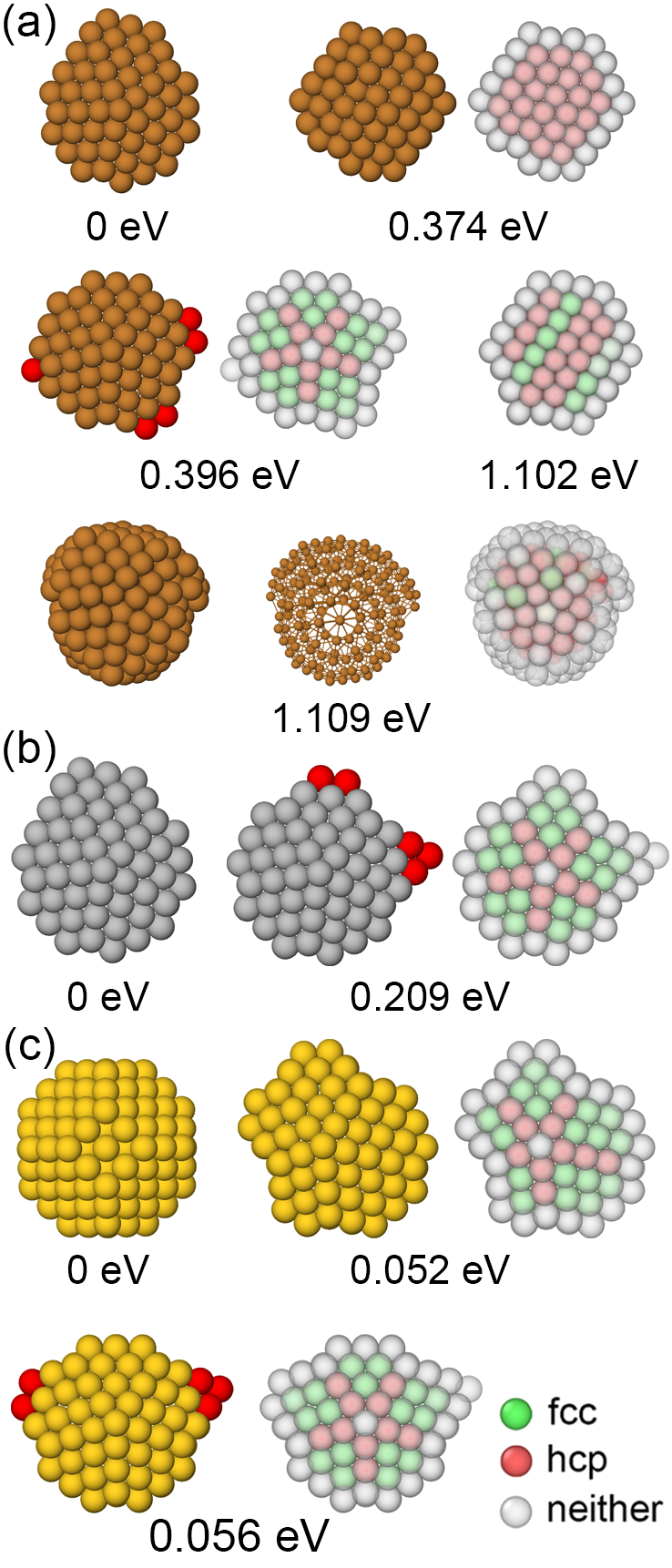}
  \caption{Structures of (a) Cu$_{201}$, (b) Ag$_{201}$, and (c) Au$_{201}$. The energy of each structure is relative to the global minimum (0 eV). The atoms in red indicate the additional nine atoms that are arranged on 192-atom Marks' decahedron to form various 201-atom decahedra.} 
  \label{fgr:structures_201}
\end{figure}

Finally, at size 201, all three systems have Dh as the dominant motif at room temperature. In Cu$_{201}$ and Ag$_{201}$, the various decahedra that are observed are all obtained by differing arrangements of nine additional atoms on magic sized 192-atom Marks decahedron. The nine additional atoms are indicated in red (see Figs. \ref{fgr:structures_201}a, b). The twins in Cu$_{201}$ have significant amount of hcp regions and are either completely hcp or consist of stacking faults. At higher temperatures, we observe icosahedra which are incomplete 309-atom icosahedra. In Au$_{201}$, Dh is the dominant motif. In this case the best Dh (second structure in Fig. \ref{fgr:structures_201}c) is different from the typical decahedra observed in Cu and Ag which are formed by adding nine atoms to the 192-atom Decahedron. Instead, the best Dh of  Au$_{201}$ is highly asymmetrical with deep reentrant grooves. However, at higher temperatures, we do observe Dh structures similar to those of Cu and Ag (third structure in Fig. \ref{fgr:structures_201}c).

\begin{figure}[!t]
\centering
  \includegraphics[width=0.4\columnwidth]{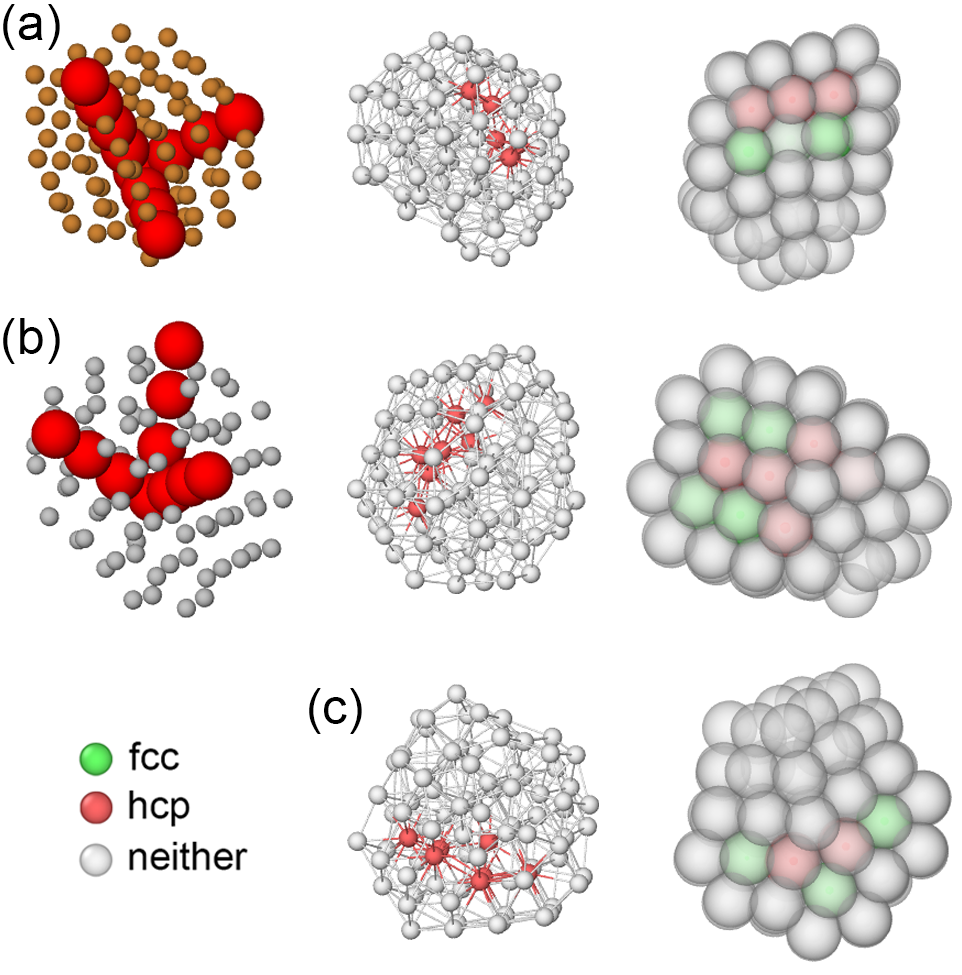}
  \caption{Mixed structures of (a) Cu$_{90}$, (b) Ag$_{90}$, and (c) Au$_{90}$. The large red atoms in (a) and (b) indicate the various decahedral axes. In (a) and (b), first image is a polydecahedron (p-Dh), second image is an icosahedron with disordered region. Third image in (a) consists of twin region and icosahedral region. The final image in (b) and (c) are mixed structures with decahedral and icosahedral regions coexisting.}
  \label{fgr:mix_90}
\end{figure}

In addition to the structures discussed above, we also observe structures that are not straightforward to categorize. We refer to these as \emph{mix} structures which occur in greater proportions at the smallest size of 90. The typical \emph{mix} structures at the size 90 are shown in Fig. \ref{fgr:mix_90}. In polydecahedron (p-Dh),\cite{polyDh2007} more than one decahedral axis is present within the same nanocluster. Examples of Cu$_{90}$ and Ag$_{90}$ p-Dh consisting of three decahedral axes are shown in the first image of Figs. \ref{fgr:mix_90}a, b. On the other hand, p-Dh are highly uncommon in Au$_{90}$. Another type of \emph{mix} structure has icosahedral region along with disordered region. All the three systems exhibit these structures (second image in Figs. \ref{fgr:mix_90}a, b and first image in \ref{fgr:mix_90}c). A third type of \emph{mix} structure occurs when local icosahedral features are observed within fcc/twin (final image in Fig. \ref{fgr:mix_90}a) or decahedron (final image in Figs. \ref{fgr:mix_90}b, c). This type of structures are mainly observed in Au and are less common in Cu and Ag clusters. The proportion of \emph{mix} structures is significantly lower at larger sizes of 147 and 201. We observe structures similar to those at the size 90 with icosahedra mixed with disordered region being more dominant. A detailed analysis of \emph{mix} structures in Au clusters has been discussed previously.\cite{settem2022AuPTMD}

\subsection{Comparison with DFT}

\begin{figure}[!t]
\centering
  \includegraphics[width=0.8\textwidth]{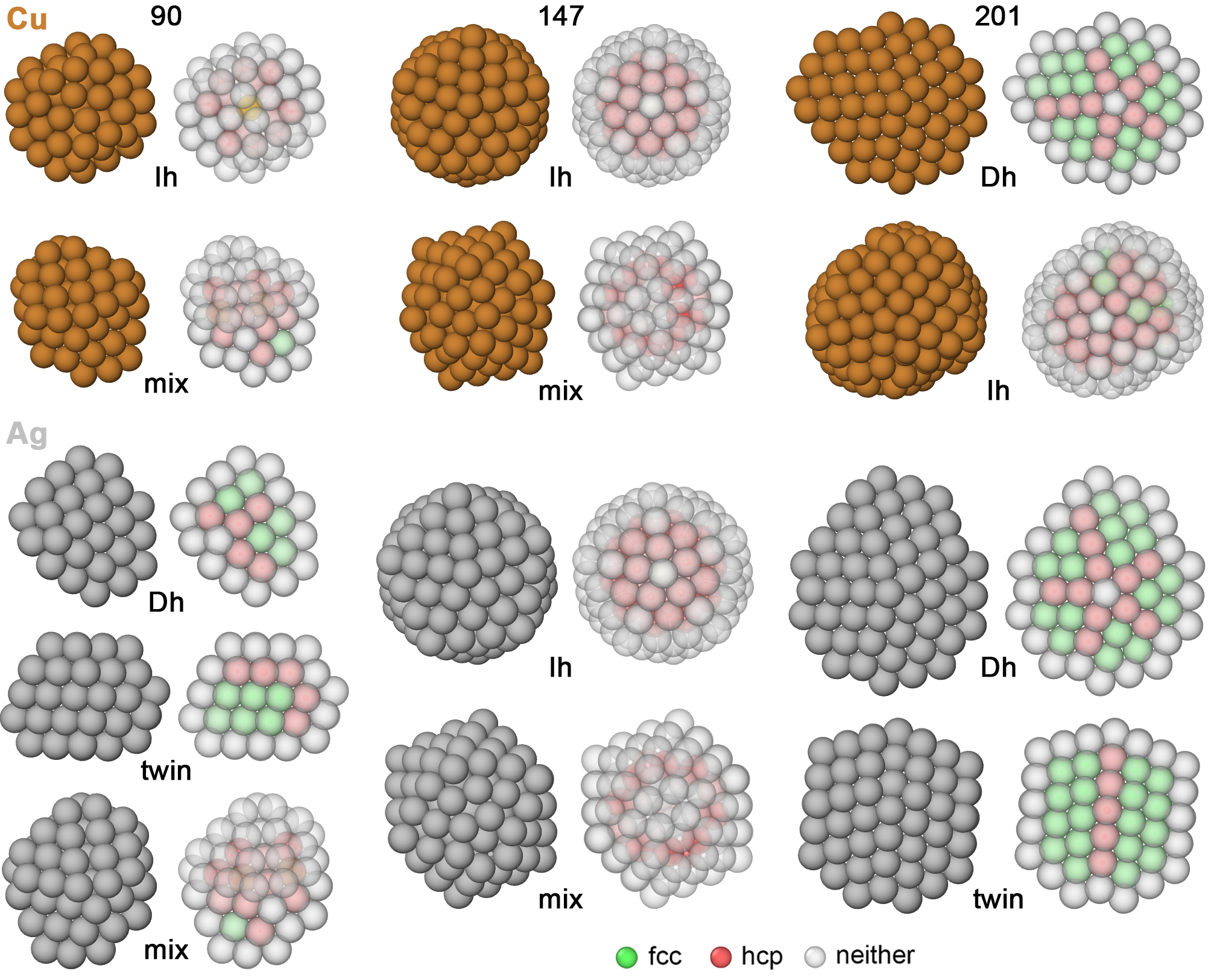}
  \caption{Cu and Ag structures used for DFT calculations.} 
  \label{fgr:dft_structures_Cu_Ag}
\end{figure}

\begin{table*}[!t]
\footnotesize
\centering
\caption{Comparison of energy differences ($\Delta$E in eV) of various motifs for Gupta potential and DFT with different exchange-correlation functional. Also, the values corresponding to embedded atom method (EAM) potentials are provided in the final column.}
{
\def\arraystretch{1.15}
\begin{tabular*}{0.88\textwidth}{@{\extracolsep{\fill}}lllllll}
\hline
\textbf{System} & \textbf{$\Delta$E} & \textbf{Gupta} & \textbf{DFT/PBE} & \textbf{DFT/LDA} & \textbf{DFT/PBEsol} & \textbf{EAM}\\
\hline
{Cu$_{90}$} & {E\textsubscript{mix}-E\textsubscript{Ih}} & {0.0828} & {1.09} & {1.10} & {1.19} & {0.2472}\\
{Cu$_{147}$} & {E\textsubscript{mix}-E\textsubscript{Ih}} & {1.5815} & {2.14} & {$-$} & {$-$} & {1.8238}\\
{Cu$_{201}$} & {E\textsubscript{Ih}-E\textsubscript{Dh}} & {0.9286} & {-0.507} & {-0.252} & {$-$} & {0.9003}\\
\hline
{Ag$_{90}$} & {E\textsubscript{mix}-E\textsubscript{Dh}} & {0.0252} & {0.159} & {0.139} & {0.149} & {-0.0043}\\
{Ag$_{90}$} & {E\textsubscript{twin}-E\textsubscript{Dh}} & {0.0319} & {-0.231} & {-0.422} & {-0.325} & {-0.1193}\\
{Ag$_{147}$} & {E\textsubscript{mix}-E\textsubscript{Ih}} & {1.0019} & {1.51} & {$-$} & {$-$} & {1.3051}\\
{Ag$_{201}$} & {E\textsubscript{twin}-E\textsubscript{Dh}} & {0.1193} & {0.609} & {$-$} & {$-$} & {0.0659}\\
\hline
{Au$_{90}$} & {E\textsubscript{twin}-E\textsubscript{fcc}} & {0.0522} & {-0.106} & {0.0761} & {0.0641} & {0.1666}\\
{Au$_{147}$} & {E\textsubscript{twin}-E\textsubscript{Dh}} & {0.0470} & {0.114} & {$-$} & {$-$} & {0.4785}\\
{Au$_{147}$} & {E\textsubscript{fcc}-E\textsubscript{Dh}} & {0.1089} & {0.616} & {$-$} & {$-$} & {0.0819}\\
{Au$_{147}$} & {E\textsubscript{mix}-E\textsubscript{Dh}} & {0.6411} & {-0.348} & {-0.330} & {-0.209} & {0.4746}\\
{Au$_{147}$} & {E\textsubscript{Ih}-E\textsubscript{Dh}} & {0.9104} & {-0.176} & {0.189} & {0.175} & {-0.3893}\\
{Au$_{147}$} & {E\textsubscript{Ih-reg}-E\textsubscript{Dh}} & {1.8649} & {2.22} & {2.07} & {1.66} & {0.1919}\\
{Au$_{201}$} & {E\textsubscript{Dh}-E\textsubscript{fcc}} & {0.0524} & {0.237} & {1.01} & {0.798} & {0.7491}\\
{Au$_{201}$} & {E\textsubscript{twin}-E\textsubscript{fcc}} & {0.0677} & {0.575} & {$-$} & {$-$} & {0.3595}\\
\hline
\end{tabular*}
}
\label{tab:gupta_vs_DFTs}
\end{table*}

\begin{figure}[!t]
\centering
  \includegraphics[width=0.8\textwidth]{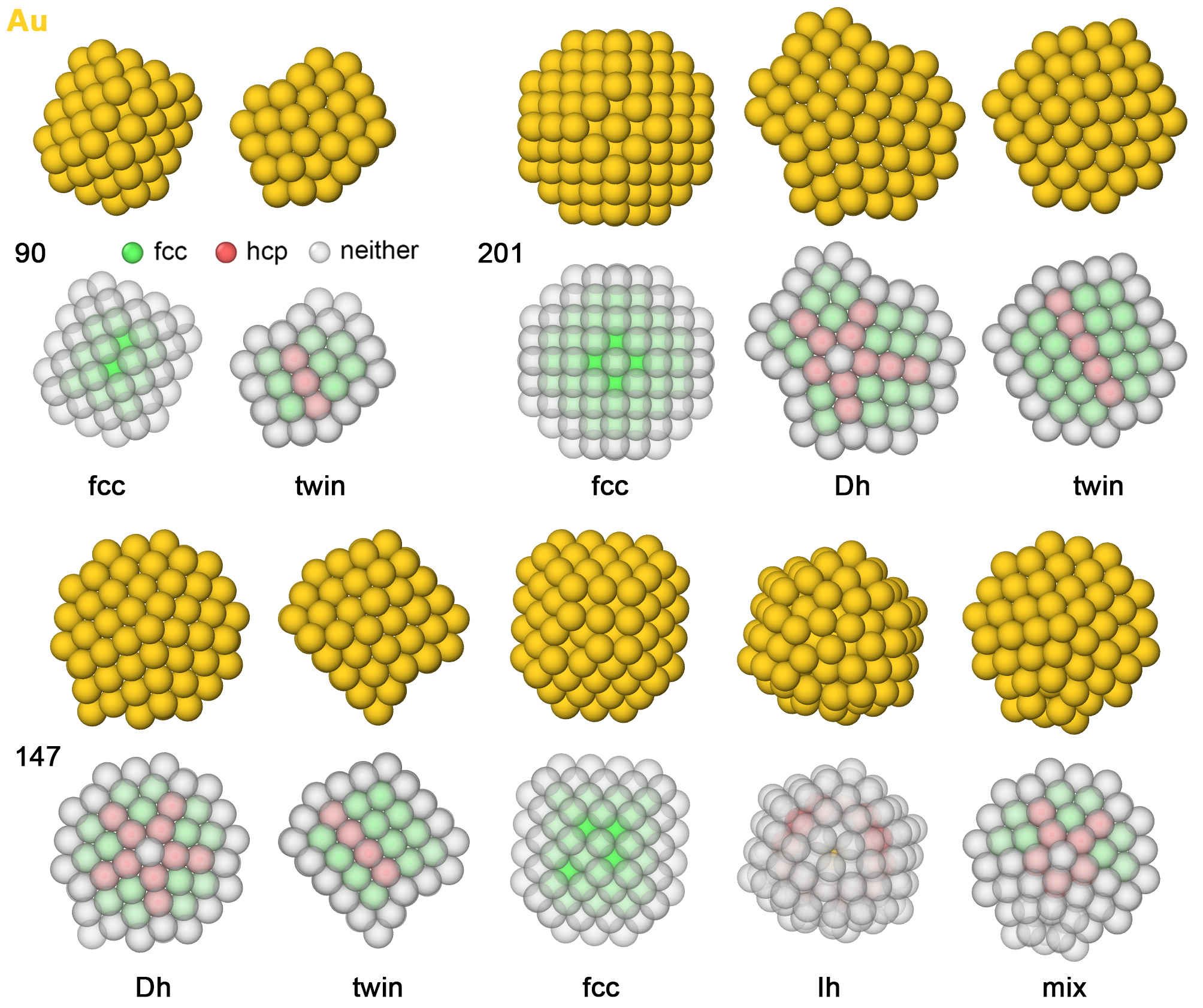}
  \caption{Au structures used for DFT calculations.} 
  \label{fgr:dft_structures_Au}
\end{figure}

\mm{The structural distributions of Cu, Ag, and Au presented so far correspond to Gupta potential which does not account for the electronic interaction between atoms. In order to assess the performance of Gupta potential, we make a comparison with DFT calculations. We used PAW pseudopotentials with three types of exchange-correlation functionals $-$ Perdew-Burke-Ernzerhof (PBE),\cite{pbeXCFunctional} local-density approximation (LDA),\cite{ldaXCFunctional} and PBE revised for solids (PBEsol).\cite{pbesolXCFunctional} }

\mm{We choose highly probable motifs (either two or more structures per each metal per each size) depending on the structural distribution. For instance, Ih and \emph{mix} are the most dominant motifs of Cu$_{90}$. In the case of Ag$_{90}$, three motifs coexist $-$ Dh, twin, and mix. Hence, we chose the lowest energy Ih, \emph{mix} for Cu$_{90}$ and Dh, twin, \emph{mix} for Ag$_{90}$. All the Cu and Ag structures used for DFT calculations are shown in Fig. \ref{fgr:dft_structures_Cu_Ag}. For a given combination of metal and size, we measure the energy difference of each structure with respect to the global minimum predicted by Gupta potential. These values are reported in Table \ref{tab:gupta_vs_DFTs} for Gupta potential, DFT/PBE, DFT/LDA, and DFT/PBEsol.}

\mm{In case of Cu$_{90}$ and Cu$_{147}$, Ih has lower energy according to both Gupta and DFT. However, for Cu$_{90}$, Ih wins by only $\sim$ 0.08 eV in comparison to $>$ 1 eV for all three DFT calculations. On the other hand, Cu$_{147}$ has a very good quantitative agreement with DFT. For Cu$_{201}$, Gupta potential predicts Dh to have lower energy than Ih in contrast to DFT. In case of Ag$_{90}$, DFT favours twin in comparison to \emph{mix} and Dh. According to DFT, the energetic ordering is E\textsubscript{twin} $<$ E\textsubscript{Dh} $<$ E\textsubscript{mix}. Gupta potential, on the other hand, predicts Dh to have the lowest energy among the three. There is a good agreement between Gupta potential and DFT for Ag$_{147}$ and Ag$_{201}$.}

\mm{Moving on to Au, the various structures used for DFT calculations are shown in Fig. \ref{fgr:dft_structures_Au}. In the case of Au$_{90}$, we observe a lack of consistency among the various exchange-correlation functionals. There is a good agreement between Gupta potential, DFT/LDA, and DFT/PBEsol with all three predicting a lower energy for fcc vs. twin. However, DFT/PBE predicts twin to be the lowest energy structure. For Au$_{147}$, we considered all the motifs (other than amorphous) given their co-existence before the melting region. Au icosahedra typically have ``Rosette'' defects. Hence, we also considered the regular closed-shell 147-atom icosahedron and refer to it as \emph{Ih-reg} in order to assess the competition between them. The energetic ordering according to Gupta potential is E\textsubscript{Ih-reg} $>$ E\textsubscript{Ih} $>$ E\textsubscript{mix} $>$ E\textsubscript{fcc} $>$ E\textsubscript{twin} $>$ E\textsubscript{Dh}. Firstly, Ih-reg has higher energy than Ih according to Gupta potential and DFT calculations confirming that Au favors defective icosahedra consisting of ``Rosette'' defects. DFT/PBE predicts Ih to have lower energy than Dh while Gupta potential, DFT/LDA, and DFT/PBEsol predict the opposite. When it comes to \emph{mix} vs. Dh, Gupta potential disagrees with DFT calculations which predict \emph{mix} to have lower energy than Dh. However, it is interesting to note that the \emph{mix} structure is indeed a Dh with local rearrangement of a few atoms near one of the reentrant grooves (see bottom of Fig. \ref{fgr:dft_structures_Au}). Hence, we believe that Dh motif will dominate also at the DFT level, in agreement with Gupta results. Finally, for Au$_{201}$, both Gupta potential and DFT predict the same energetic ordering: E\textsubscript{twin} $>$ E\textsubscript{Dh} $>$ E\textsubscript{fcc}. However, Gupta potential has lower energy difference compared to DFT. As a result, we anticipate that the solid-solid transformation from fcc $\rightarrow$ Dh will be delayed to occur at higher temperature than predicted by Gupta potential.}

\mm{Overall, we observe the following trends. At the size 147, Gupta potential performs fairly well, the more so for Cu$_{147}$ and Ag$_{147}$ which exhibit excellent quantitative agreement between Gupta potential and DFT. In the case of Au$_{147}$, Gupta potential does a good job. Firstly, it predicts that defective icosahedra are preferred with surface ``rosettes''. Secondly, Ih has higher energy than Dh and \emph{mix} according to both Gupta and DFT. The only difference is \emph{mix}, which is a distorted Dh with local rearrangement near the reentrant groove, is energetically preferred over Dh at the DFT level. At size 90, there is a qualitative agreement between Gupta potential and DFT for Cu, but not for Ag and Au. In the case of Ag$_{90}$, twin is preferred at the DFT level, while Dh is preferred according to Gupta potential. In Au$_{90}$, there is internal disagreement among DFT exchange-correlation functionals. However, given the very small energy difference (absolute values are about 0.1 eV or lower), we expect a similar competition between fcc and twin as observed with Gupta potential. Finally, at size 201, both Ag and Au exhibit a qualitative agreement with DFT (although they underestimate the energy differences). In the case of Cu$_{201}$, Ih is preferred at the DFT level as opposed to Dh according to Gupta potential.}

\mm{Finally, in order to understand how the embedded atom method (EAM) pair potential model performs in comparison to Gupta potential, we calculated the energy differences using EAM potentials for Cu,\cite{potCuAg} Ag,\cite{potCuAg} and Au.\cite{potAuEAM} The results are reported in the final column of Table \ref{tab:gupta_vs_DFTs}. In the case of Cu, Gupta and EAM exhibit similar performance. On the other hand, EAM seems to perform marginally better in the case of Ag. According to EAM, Ag$_{90}$ predicts twin to have lower energy compared to \emph{mix} and Dh similar to DFT. In the case of Au, EAM performs poorly in comparison to Gupta. The major drawback with EAM is that it predicts Ih to be the lowest energy structure for Au$_{147}$.}

\textcolor{black}{Based on these results, we find that model potentials are still a good guide to select the main structural motifs and for discussing general trends between metals, but in some cases they fail to select the lowest-energy motifs in agreement with DFT. We note however that there is a case, Au$_{90}$, where there is qualitative disagreement even between different types of DFT calculations. Moreover, in general there are quantitative discrepancies between the different exchange-correlation functionals, which would make it difficult to assign precise temperature-dependent isomer probabilities even at the DFT level.}

\section{Conclusions}
In this work, we have applied a computational framework that we proposed recently\cite{settem2022AuPTMD} to study the size- and system-dependent structural distributions of Cu, Ag, and Au nanoclusters. In this method, we combine harmonic superposition approximation (HSA) and parallel tempering molecular dynamics (PTMD) in a complementary manner and calculate the structures of metal nanoclusters in the entire temperature range from 0 K to melting. We considered three cluster sizes $-$ 90, 147, and 201 in the range 1 to 2 nm of which 147 and 201 are ``magic'' sizes.

To begin with, ``magic'' sizes are not necessarily ``magic'' in that the global minimum is not always the ideal geometrical motif at that size. Perfect icosahedron and truncated octahedron are the ideal geometries at the sizes 147 and 201, respectively. However, only in three out of six cases (Cu$_{147}$, Ag$_{147}$, Au$_{201}$) the global minimum corresponds to the ideal geometrical structure. The global minima of Au$_{147}$, Cu$_{201}$, and Ag$_{201}$ are all Marks decahedra. At size 90, all the three systems have a different global minimum: icosahedron for Cu$_{90}$, decahedron for Ag$_{90}$, and fcc for Au$_{90}$.

The structural changes in these systems can be categorised broadly into three groups: type-(i) global minimum is also the dominant motif at finite temperatures up to melting; type-(ii) solid-solid transformations lead to a completely different motif; type-(iii) solid-solid transitions lead to a co-existence of two or more motifs. The majority of the cases belong to the second and third groups, which include \bb{Cu$_{90}$, Cu$_{201}$, Ag$_{90}$, Au$_{90}$, Au$_{147}$, and Au$_{201}$}.

Icosahedra are extremely dominant with almost 100\% abundance in Cu$_{147}$ and Ag$_{147}$ right up to melting. Similarly, decahedra are the dominant motif in Au$_{201}$ up to melting. In the cases of Cu$_{90}$ and Au$_{90}$, we find find significant proportion of \emph{mix} structures close to melting. Although decahedra are dominant in Cu$_{201}$, we find significant amount of icosahedra beyond 400 K. Finally, in Au$_{147}$, the proportion of Dh decreases gradually and we find small amounts of twin, fcc, Ih and \emph{mix} structures co-existing at higher temperatures.

In contrast, Ag$_{90}$ and Au$_{201}$ undergo solid-solid transformations. Ag$_{90}$ exhibits a partial transformation Dh $\rightarrow$ Dh $+$ twin $+$ \emph{mix} between 100 K to 150 K. Beyond 150 K, the proportion of Dh, twin,and \emph{mix} structures remains approximately constant up to 450 K indicating a co-existence of multiple motifs. In the case of Au$_{201}$, fcc transforms to Dh below 200 K resulting in almost 100\% Dh at room temperature which remains dominant up to melting. In both the instances, the solid-solid transformation occurs well below the room temperature ($<$ 200 K). As a result, it is non-trivial to predict the finite-temperature structures from the global minimum alone.

We also observed system specific differences across the three metals. Cu has a stronger preference for icosahedral structures. This is evident from almost 100\% abundance at the sizes 90 and 147. While, at size 201, a significant amount of icosahedra are observed above 400 K which peaks around 700 K with $\sim$ 33\%. In the case of Ag, icosahedra are mainly observed at the ``magic'' size of 147 where they occur with almost 100\% abundance. Au on the other hand disfavors icosahedra, with icosahedra observed mainly at the size 147 in small proportions beyond 400 K. Another interesting feature is the gradual change in the nature of ``rosette'' defects in icosahedra at the size 147 from Cu to Au. ``Rosette''defects are completely absent in Cu, but appear at higher temperatures in Ag. However, typically, almost all the icosahedra in Au have ``rosette'' defects. In contrast to Cu and Ag, decahedra in Au have deeper reentrant grooves.

\mm{Finally, a comparison of the performance of Gupta potential with DFT reveals few limitations of interatomic pair potentials. We observe a good agreement between Gupta potential and DFT at the size 147. In other cases (Cu$_{90}$, Ag$_{90}$, Ag$_{201}$, and Au$_{201}$), the energetic ordering of the considered motifs is same according to both Gupta potential and DFT, with Gupta energy differences being underestimated. In the case of Au$_{90}$, Gupta potential agrees with DFT/LDA, DFT/PBEsol but not with DFT/PBE. Finally, Gupta potential fares poorly in the case of Cu$_{201}$ since it predicts Dh to prevail over Ih. However, according to DFT, Ih should prevail over Dh. Notwithstanding these limitations, interatomic pair potentials remain indispensable since the wide exporation of the energy landscape of metal nanoclusters at the DFT level is simply not feasible. It is instructive to first obtain the structural distributions using interatomic pair potentials, e.g., Gupta as done in the current work, followed by DFT calculations to understand the limitations of the structural distributions. For instance, we observed that Gupta potential predicts a complete solid-solid transformation from fcc $\rightarrow$ Dh below room temperature for Au$_{201}$. However, the energy difference between Dh and fcc is lower than predicted by DFT (0.0524 eV for Gupta potential vs. $>$ 0.2 eV for DFT). Based on this information, it can be inferred that the transformation from fcc $\rightarrow$ Dh may occur at higher temperature than predicted by Gupta potential. A further check of another model potential, EAM, shows an overall performance of the same quality of the Gupta potential, with a better agreement with DFT for Ag$_{90}$ and a poorer performance for Au clusters.}

Our method can be easily applied to any size and system for which reasonable models for atomic interactions are available. As a result, this method enables one to estimate the equilibrium proportion of various geometrical motifs as a function of temperature which can then be used to compare with the experimentally obtained structural distribution.\cite{loffreda2021Ag309TEM,wells2015AuImagingACFraction,foster2018AuImagingACFraction} This allows one to verify if the experimentally observed structures are in equilibrium or kinetically trapped metastable structures.

\section*{Supplementary Material}
Supplementary material contains Parameters of HSA, PTMD; Structural distribution of Au nanoclusters.

\begin{acknowledgments}
This work has been supported by the project ``Understanding and Tuning FRiction through nanOstructure Manipulation (UTFROM)'' funded by MIUR  Progetti di Ricerca di Rilevante Interesse Nazionale (PRIN) Bando 2017 - grant 20178PZCB5. M.S. and A.G. acknowledge financial support from MIUR "Framework per l'Attrazione e il Rafforzamento delle Eccellenze per la Ricerca in Italia (FARE)” scheme, grant SERENA n. R18XYKRW7J. R.F. acknowledges the Progetto di Eccellenza of the Physics Department of the University of Genoa for financial and the International Research Network Nanoalloys of CNRS for networking support. The authors acknowledge PRACE for awarding us access to Marconi100 at CINECA, Italy.
\end{acknowledgments}

\section*{Data Availability Statement}
The data that support the findings of this study are available from the corresponding author upon reasonable request.

\section*{Author Declarations}
The authors have no conflicts to disclose.

\bibliography{aipsamp}

\begin{thebibliography}{91}%
\makeatletter
\providecommand \@ifxundefined [1]{%
 \@ifx{#1\undefined}
}%
\providecommand \@ifnum [1]{%
 \ifnum #1\expandafter \@firstoftwo
 \else \expandafter \@secondoftwo
 \fi
}%
\providecommand \@ifx [1]{%
 \ifx #1\expandafter \@firstoftwo
 \else \expandafter \@secondoftwo
 \fi
}%
\providecommand \natexlab [1]{#1}%
\providecommand \enquote  [1]{``#1''}%
\providecommand \bibnamefont  [1]{#1}%
\providecommand \bibfnamefont [1]{#1}%
\providecommand \citenamefont [1]{#1}%
\providecommand \href@noop [0]{\@secondoftwo}%
\providecommand \href [0]{\begingroup \@sanitize@url \@href}%
\providecommand \@href[1]{\@@startlink{#1}\@@href}%
\providecommand \@@href[1]{\endgroup#1\@@endlink}%
\providecommand \@sanitize@url [0]{\catcode `\\12\catcode `\$12\catcode
  `\&12\catcode `\#12\catcode `\^12\catcode `\_12\catcode `\%12\relax}%
\providecommand \@@startlink[1]{}%
\providecommand \@@endlink[0]{}%
\providecommand \url  [0]{\begingroup\@sanitize@url \@url }%
\providecommand \@url [1]{\endgroup\@href {#1}{\urlprefix }}%
\providecommand \urlprefix  [0]{URL }%
\providecommand \Eprint [0]{\href }%
\providecommand \doibase [0]{http://dx.doi.org/}%
\providecommand \selectlanguage [0]{\@gobble}%
\providecommand \bibinfo  [0]{\@secondoftwo}%
\providecommand \bibfield  [0]{\@secondoftwo}%
\providecommand \translation [1]{[#1]}%
\providecommand \BibitemOpen [0]{}%
\providecommand \bibitemStop [0]{}%
\providecommand \bibitemNoStop [0]{.\EOS\space}%
\providecommand \EOS [0]{\spacefactor3000\relax}%
\providecommand \BibitemShut  [1]{\csname bibitem#1\endcsname}%
\let\auto@bib@innerbib\@empty
\bibitem [{\citenamefont {Baletto}\ and\ \citenamefont
  {Ferrando}(2005)}]{baletto2005rev}%
  \BibitemOpen
  \bibfield  {author} {\bibinfo {author} {\bibfnamefont {F.}~\bibnamefont
  {Baletto}}\ and\ \bibinfo {author} {\bibfnamefont {R.}~\bibnamefont
  {Ferrando}},\ }\href {\doibase 10.1103/RevModPhys.77.371} {\bibfield
  {journal} {\bibinfo  {journal} {Rev. Mod. Phys.}\ }\textbf {\bibinfo {volume}
  {77}},\ \bibinfo {pages} {371} (\bibinfo {year} {2005})}\BibitemShut
  {NoStop}%
\bibitem [{\citenamefont {Baletto}\ \emph {et~al.}(2002)\citenamefont
  {Baletto}, \citenamefont {Ferrando}, \citenamefont {Fortunelli},
  \citenamefont {Montalenti},\ and\ \citenamefont
  {Mottet}}]{baletto2002potParams}%
  \BibitemOpen
  \bibfield  {author} {\bibinfo {author} {\bibfnamefont {F.}~\bibnamefont
  {Baletto}}, \bibinfo {author} {\bibfnamefont {R.}~\bibnamefont {Ferrando}},
  \bibinfo {author} {\bibfnamefont {A.}~\bibnamefont {Fortunelli}}, \bibinfo
  {author} {\bibfnamefont {F.}~\bibnamefont {Montalenti}}, \ and\ \bibinfo
  {author} {\bibfnamefont {C.}~\bibnamefont {Mottet}},\ }\href {\doibase
  10.1063/1.1448484} {\bibfield  {journal} {\bibinfo  {journal} {J. Chem.
  Phys.}\ }\textbf {\bibinfo {volume} {116}},\ \bibinfo {pages} {3856}
  (\bibinfo {year} {2002})}\BibitemShut {NoStop}%
\bibitem [{\citenamefont {Rahm}\ and\ \citenamefont
  {Erhart}(2017)}]{rahm2017coExist}%
  \BibitemOpen
  \bibfield  {author} {\bibinfo {author} {\bibfnamefont {J.~M.}\ \bibnamefont
  {Rahm}}\ and\ \bibinfo {author} {\bibfnamefont {P.}~\bibnamefont {Erhart}},\
  }\href {\doibase 10.1021/acs.nanolett.7b02761} {\bibfield  {journal}
  {\bibinfo  {journal} {Nano Lett.}\ }\textbf {\bibinfo {volume} {17}},\
  \bibinfo {pages} {5775} (\bibinfo {year} {2017})}\BibitemShut {NoStop}%
\bibitem [{\citenamefont {Nelli}, \citenamefont {Roncaglia},\ and\
  \citenamefont {Minnai}(2023)}]{Nelli2023apx}%
  \BibitemOpen
  \bibfield  {author} {\bibinfo {author} {\bibfnamefont {D.}~\bibnamefont
  {Nelli}}, \bibinfo {author} {\bibfnamefont {C.}~\bibnamefont {Roncaglia}}, \
  and\ \bibinfo {author} {\bibfnamefont {C.}~\bibnamefont {Minnai}},\ }\href
  {\doibase 10.1080/23746149.2022.2127330} {\bibfield  {journal} {\bibinfo
  {journal} {Advances in Physics: X}\ }\textbf {\bibinfo {volume} {8}},\
  \bibinfo {pages} {2127330} (\bibinfo {year} {2023})}\BibitemShut {NoStop}%
\bibitem [{\citenamefont {Michaelian}, \citenamefont {Rendon},\ and\
  \citenamefont {Garzon}(1999)}]{michaelian1999AuRelSRAmor}%
  \BibitemOpen
  \bibfield  {author} {\bibinfo {author} {\bibfnamefont {K.}~\bibnamefont
  {Michaelian}}, \bibinfo {author} {\bibfnamefont {N.}~\bibnamefont {Rendon}},
  \ and\ \bibinfo {author} {\bibfnamefont {I.~L.}\ \bibnamefont {Garzon}},\
  }\href {\doibase 10.1103/PhysRevB.60.2000} {\bibfield  {journal} {\bibinfo
  {journal} {Phys. Rev. B}\ }\textbf {\bibinfo {volume} {60}},\ \bibinfo
  {pages} {2000} (\bibinfo {year} {1999})}\BibitemShut {NoStop}%
\bibitem [{\citenamefont {Grigoryan}, \citenamefont {Alamanova},\ and\
  \citenamefont {Springborg}(2005)}]{grigoryan2005NiCuAu60}%
  \BibitemOpen
  \bibfield  {author} {\bibinfo {author} {\bibfnamefont {V.~G.}\ \bibnamefont
  {Grigoryan}}, \bibinfo {author} {\bibfnamefont {D.}~\bibnamefont
  {Alamanova}}, \ and\ \bibinfo {author} {\bibfnamefont {M.}~\bibnamefont
  {Springborg}},\ }\href {\doibase 10.1140/epjd/e2005-00141-6} {\bibfield
  {journal} {\bibinfo  {journal} {Eur. Phys. J. D}\ }\textbf {\bibinfo {volume}
  {34}},\ \bibinfo {pages} {187} (\bibinfo {year} {2005})}\BibitemShut
  {NoStop}%
\bibitem [{\citenamefont {Shao}, \citenamefont {Liu},\ and\ \citenamefont
  {Cai}(2005)}]{shao2005Ag80}%
  \BibitemOpen
  \bibfield  {author} {\bibinfo {author} {\bibfnamefont {X.}~\bibnamefont
  {Shao}}, \bibinfo {author} {\bibfnamefont {X.}~\bibnamefont {Liu}}, \ and\
  \bibinfo {author} {\bibfnamefont {W.}~\bibnamefont {Cai}},\ }\href {\doibase
  10.1021/ct049865j} {\bibfield  {journal} {\bibinfo  {journal} {J. Chem.
  Theory Comput.}\ }\textbf {\bibinfo {volume} {1}},\ \bibinfo {pages} {762}
  (\bibinfo {year} {2005})}\BibitemShut {NoStop}%
\bibitem [{\citenamefont {Apra}, \citenamefont {Ferrando},\ and\ \citenamefont
  {Fortunelli}(2006)}]{apra2006AuRelSRAmor}%
  \BibitemOpen
  \bibfield  {author} {\bibinfo {author} {\bibfnamefont {E.}~\bibnamefont
  {Apra}}, \bibinfo {author} {\bibfnamefont {R.}~\bibnamefont {Ferrando}}, \
  and\ \bibinfo {author} {\bibfnamefont {A.}~\bibnamefont {Fortunelli}},\
  }\href {\doibase 10.1103/PhysRevB.73.205414} {\bibfield  {journal} {\bibinfo
  {journal} {Phys. Rev. B}\ }\textbf {\bibinfo {volume} {73}},\ \bibinfo
  {pages} {205414} (\bibinfo {year} {2006})}\BibitemShut {NoStop}%
\bibitem [{\citenamefont {Grigoryan}, \citenamefont {Alamanova},\ and\
  \citenamefont {Springborg}(2006)}]{grigoryan2006Cu150}%
  \BibitemOpen
  \bibfield  {author} {\bibinfo {author} {\bibfnamefont {V.~G.}\ \bibnamefont
  {Grigoryan}}, \bibinfo {author} {\bibfnamefont {D.}~\bibnamefont
  {Alamanova}}, \ and\ \bibinfo {author} {\bibfnamefont {M.}~\bibnamefont
  {Springborg}},\ }\href {\doibase 10.1103/PhysRevB.73.115415} {\bibfield
  {journal} {\bibinfo  {journal} {Phys. Rev. B}\ }\textbf {\bibinfo {volume}
  {73}},\ \bibinfo {pages} {115415} (\bibinfo {year} {2006})}\BibitemShut
  {NoStop}%
\bibitem [{\citenamefont {Yang}, \citenamefont {Cai},\ and\ \citenamefont
  {Shao}(2007)}]{yang2007Ag160}%
  \BibitemOpen
  \bibfield  {author} {\bibinfo {author} {\bibfnamefont {X.}~\bibnamefont
  {Yang}}, \bibinfo {author} {\bibfnamefont {W.}~\bibnamefont {Cai}}, \ and\
  \bibinfo {author} {\bibfnamefont {X.}~\bibnamefont {Shao}},\ }\href {\doibase
  10.1021/jp0711895} {\bibfield  {journal} {\bibinfo  {journal} {J. Phys. Chem.
  A}\ }\textbf {\bibinfo {volume} {111}},\ \bibinfo {pages} {5048} (\bibinfo
  {year} {2007})}\BibitemShut {NoStop}%
\bibitem [{\citenamefont {Alamanova}, \citenamefont {Grigoryan},\ and\
  \citenamefont {Springborg}(2007)}]{alamanova2007Ag150}%
  \BibitemOpen
  \bibfield  {author} {\bibinfo {author} {\bibfnamefont {D.}~\bibnamefont
  {Alamanova}}, \bibinfo {author} {\bibfnamefont {V.~G.}\ \bibnamefont
  {Grigoryan}}, \ and\ \bibinfo {author} {\bibfnamefont {M.}~\bibnamefont
  {Springborg}},\ }\href {\doibase 10.1021/jp0717342} {\bibfield  {journal}
  {\bibinfo  {journal} {J. Phys. Chem. C}\ }\textbf {\bibinfo {volume} {111}},\
  \bibinfo {pages} {12577} (\bibinfo {year} {2007})}\BibitemShut {NoStop}%
\bibitem [{\citenamefont {Angulo}\ and\ \citenamefont
  {Noguez}(2008)}]{angulo2008Ag}%
  \BibitemOpen
  \bibfield  {author} {\bibinfo {author} {\bibfnamefont {A.~M.}\ \bibnamefont
  {Angulo}}\ and\ \bibinfo {author} {\bibfnamefont {C.}~\bibnamefont
  {Noguez}},\ }\href {\doibase 10.1021/jp801545x} {\bibfield  {journal}
  {\bibinfo  {journal} {J. Phys. Chem. A}\ }\textbf {\bibinfo {volume} {112}},\
  \bibinfo {pages} {5834} (\bibinfo {year} {2008})}\BibitemShut {NoStop}%
\bibitem [{\citenamefont {Itoh}\ \emph {et~al.}(2009)\citenamefont {Itoh},
  \citenamefont {Kumar}, \citenamefont {Adschiri},\ and\ \citenamefont
  {Kawazoe}}]{itoh2009CuAg75}%
  \BibitemOpen
  \bibfield  {author} {\bibinfo {author} {\bibfnamefont {M.}~\bibnamefont
  {Itoh}}, \bibinfo {author} {\bibfnamefont {V.}~\bibnamefont {Kumar}},
  \bibinfo {author} {\bibfnamefont {T.}~\bibnamefont {Adschiri}}, \ and\
  \bibinfo {author} {\bibfnamefont {Y.}~\bibnamefont {Kawazoe}},\ }\href
  {\doibase 10.1063/1.3187934} {\bibfield  {journal} {\bibinfo  {journal} {J.
  Chem. Phys.}\ }\textbf {\bibinfo {volume} {131}},\ \bibinfo {pages} {174510}
  (\bibinfo {year} {2009})}\BibitemShut {NoStop}%
\bibitem [{\citenamefont {Bao}\ \emph {et~al.}(2009)\citenamefont {Bao},
  \citenamefont {Goedecker}, \citenamefont {Koga}, \citenamefont {Lançon},\
  and\ \citenamefont {Neelov}}]{bao2009AuBHMC}%
  \BibitemOpen
  \bibfield  {author} {\bibinfo {author} {\bibfnamefont {K.}~\bibnamefont
  {Bao}}, \bibinfo {author} {\bibfnamefont {S.}~\bibnamefont {Goedecker}},
  \bibinfo {author} {\bibfnamefont {K.}~\bibnamefont {Koga}}, \bibinfo {author}
  {\bibfnamefont {F.}~\bibnamefont {Lançon}}, \ and\ \bibinfo {author}
  {\bibfnamefont {A.}~\bibnamefont {Neelov}},\ }\href {\doibase
  10.1103/PhysRevB.79.041405} {\bibfield  {journal} {\bibinfo  {journal} {Phys.
  Rev. B}\ }\textbf {\bibinfo {volume} {79}},\ \bibinfo {pages} {041405(R)}
  (\bibinfo {year} {2009})}\BibitemShut {NoStop}%
\bibitem [{\citenamefont {Huang}, \citenamefont {Lai},\ and\ \citenamefont
  {Xu}(2011)}]{huang2011Ag141to310}%
  \BibitemOpen
  \bibfield  {author} {\bibinfo {author} {\bibfnamefont {W.}~\bibnamefont
  {Huang}}, \bibinfo {author} {\bibfnamefont {X.}~\bibnamefont {Lai}}, \ and\
  \bibinfo {author} {\bibfnamefont {R.}~\bibnamefont {Xu}},\ }\href {\doibase
  10.1016/j.cplett.2011.03.070} {\bibfield  {journal} {\bibinfo  {journal}
  {Chem. Phys. Lett.}\ }\textbf {\bibinfo {volume} {507}},\ \bibinfo {pages}
  {199} (\bibinfo {year} {2011})}\BibitemShut {NoStop}%
\bibitem [{\citenamefont {Chen}\ \emph {et~al.}(2013)\citenamefont {Chen},
  \citenamefont {Dyer}, \citenamefont {Li},\ and\ \citenamefont
  {Dixon}}]{chen2013Ag99}%
  \BibitemOpen
  \bibfield  {author} {\bibinfo {author} {\bibfnamefont {M.}~\bibnamefont
  {Chen}}, \bibinfo {author} {\bibfnamefont {J.~E.}\ \bibnamefont {Dyer}},
  \bibinfo {author} {\bibfnamefont {K.}~\bibnamefont {Li}}, \ and\ \bibinfo
  {author} {\bibfnamefont {D.~A.}\ \bibnamefont {Dixon}},\ }\href {\doibase
  10.1021/jp404493w} {\bibfield  {journal} {\bibinfo  {journal} {J. Phys. Chem.
  A}\ }\textbf {\bibinfo {volume} {117}},\ \bibinfo {pages} {8298} (\bibinfo
  {year} {2013})}\BibitemShut {NoStop}%
\bibitem [{\citenamefont {Grigoryan}\ \emph {et~al.}(2013)\citenamefont
  {Grigoryan}, \citenamefont {Springborg}, \citenamefont {Minassian},\ and\
  \citenamefont {Melikyan}}]{grigoryan2013AgCu150}%
  \BibitemOpen
  \bibfield  {author} {\bibinfo {author} {\bibfnamefont {V.~G.}\ \bibnamefont
  {Grigoryan}}, \bibinfo {author} {\bibfnamefont {M.}~\bibnamefont
  {Springborg}}, \bibinfo {author} {\bibfnamefont {H.}~\bibnamefont
  {Minassian}}, \ and\ \bibinfo {author} {\bibfnamefont {A.}~\bibnamefont
  {Melikyan}},\ }\href {\doibase 10.1016/j.comptc.2013.07.022} {\bibfield
  {journal} {\bibinfo  {journal} {Comput. Theor. Chem.}\ }\textbf {\bibinfo
  {volume} {1021}},\ \bibinfo {pages} {197} (\bibinfo {year}
  {2013})}\BibitemShut {NoStop}%
\bibitem [{\citenamefont {Schebarchov}, \citenamefont {Baletto},\ and\
  \citenamefont {Wales}(2018)}]{schebarchov2018AuHSA}%
  \BibitemOpen
  \bibfield  {author} {\bibinfo {author} {\bibfnamefont {D.}~\bibnamefont
  {Schebarchov}}, \bibinfo {author} {\bibfnamefont {F.}~\bibnamefont
  {Baletto}}, \ and\ \bibinfo {author} {\bibfnamefont {D.~J.}\ \bibnamefont
  {Wales}},\ }\href {\doibase 10.1039/c7nr07123j} {\bibfield  {journal}
  {\bibinfo  {journal} {Nanoscale}\ }\textbf {\bibinfo {volume} {10}},\
  \bibinfo {pages} {2004} (\bibinfo {year} {2018})}\BibitemShut {NoStop}%
\bibitem [{\citenamefont {Settem}, \citenamefont {Ferrando},\ and\
  \citenamefont {Giacomello}(2022)}]{settem2022AuPTMD}%
  \BibitemOpen
  \bibfield  {author} {\bibinfo {author} {\bibfnamefont {M.}~\bibnamefont
  {Settem}}, \bibinfo {author} {\bibfnamefont {R.}~\bibnamefont {Ferrando}}, \
  and\ \bibinfo {author} {\bibfnamefont {A.}~\bibnamefont {Giacomello}},\
  }\href {\doibase 10.1039/D1NR05078H} {\bibfield  {journal} {\bibinfo
  {journal} {Nanoscale}\ }\textbf {\bibinfo {volume} {14}},\ \bibinfo {pages}
  {939} (\bibinfo {year} {2022})}\BibitemShut {NoStop}%
\bibitem [{\citenamefont {Franke}, \citenamefont {Hilf},\ and\ \citenamefont
  {Borrmann}(1993)}]{hsa1993}%
  \BibitemOpen
  \bibfield  {author} {\bibinfo {author} {\bibfnamefont {G.}~\bibnamefont
  {Franke}}, \bibinfo {author} {\bibfnamefont {E.~R.}\ \bibnamefont {Hilf}}, \
  and\ \bibinfo {author} {\bibfnamefont {P.}~\bibnamefont {Borrmann}},\ }\href
  {\doibase 10.1063/1.464070} {\bibfield  {journal} {\bibinfo  {journal} {J.
  Chern. Phys.}\ }\textbf {\bibinfo {volume} {98}},\ \bibinfo {pages} {3496}
  (\bibinfo {year} {1993})}\BibitemShut {NoStop}%
\bibitem [{\citenamefont {Calvo}, \citenamefont {Doye},\ and\ \citenamefont
  {Wales}(2002)}]{calvo2002}%
  \BibitemOpen
  \bibfield  {author} {\bibinfo {author} {\bibfnamefont {F.}~\bibnamefont
  {Calvo}}, \bibinfo {author} {\bibfnamefont {J.~P.~K.}\ \bibnamefont {Doye}},
  \ and\ \bibinfo {author} {\bibfnamefont {D.~J.}\ \bibnamefont {Wales}},\
  }\href {\doibase 10.1016/S0009-2614(02)01550-6} {\bibfield  {journal}
  {\bibinfo  {journal} {Chem. Phys. Lett.}\ }\textbf {\bibinfo {volume}
  {366}},\ \bibinfo {pages} {176} (\bibinfo {year} {2002})}\BibitemShut
  {NoStop}%
\bibitem [{\citenamefont {Doye}\ and\ \citenamefont
  {Calvo}(2001)}]{doye2001hsaLJ}%
  \BibitemOpen
  \bibfield  {author} {\bibinfo {author} {\bibfnamefont {J.~P.~K.}\
  \bibnamefont {Doye}}\ and\ \bibinfo {author} {\bibfnamefont {F.}~\bibnamefont
  {Calvo}},\ }\href {\doibase 10.1103/PhysRevLett.86.3570} {\bibfield
  {journal} {\bibinfo  {journal} {Phys. Rev. Lett.}\ }\textbf {\bibinfo
  {volume} {86}},\ \bibinfo {pages} {3570} (\bibinfo {year}
  {2001})}\BibitemShut {NoStop}%
\bibitem [{\citenamefont {Doye}\ and\ \citenamefont
  {Calvo}(2002)}]{doye2002hsaLJ}%
  \BibitemOpen
  \bibfield  {author} {\bibinfo {author} {\bibfnamefont {J.~P.~K.}\
  \bibnamefont {Doye}}\ and\ \bibinfo {author} {\bibfnamefont {F.}~\bibnamefont
  {Calvo}},\ }\href {\doibase 10.1063/1.1469616} {\bibfield  {journal}
  {\bibinfo  {journal} {J. Chem. Phys.}\ }\textbf {\bibinfo {volume} {116}},\
  \bibinfo {pages} {8307} (\bibinfo {year} {2002})}\BibitemShut {NoStop}%
\bibitem [{\citenamefont {Mandelshtam}\ and\ \citenamefont
  {Frantsuzov}(2006)}]{mandelshtam2006hsaLJ}%
  \BibitemOpen
  \bibfield  {author} {\bibinfo {author} {\bibfnamefont {V.~A.}\ \bibnamefont
  {Mandelshtam}}\ and\ \bibinfo {author} {\bibfnamefont {P.~A.}\ \bibnamefont
  {Frantsuzov}},\ }\href {\doibase 10.1063/1.2202312} {\bibfield  {journal}
  {\bibinfo  {journal} {J. Chem. Phys.}\ }\textbf {\bibinfo {volume} {124}},\
  \bibinfo {pages} {204511} (\bibinfo {year} {2006})}\BibitemShut {NoStop}%
\bibitem [{\citenamefont {Sharapov}\ and\ \citenamefont
  {Mandelshtam}(2007)}]{sharapov2007hsaLJ}%
  \BibitemOpen
  \bibfield  {author} {\bibinfo {author} {\bibfnamefont {V.~A.}\ \bibnamefont
  {Sharapov}}\ and\ \bibinfo {author} {\bibfnamefont {V.~A.}\ \bibnamefont
  {Mandelshtam}},\ }\href {\doibase 10.1021/jp072929c} {\bibfield  {journal}
  {\bibinfo  {journal} {J. Phys. Chem. A}\ }\textbf {\bibinfo {volume} {111}},\
  \bibinfo {pages} {10284} (\bibinfo {year} {2007})}\BibitemShut {NoStop}%
\bibitem [{\citenamefont {Grigoryan}\ and\ \citenamefont
  {Springborg}(2019)}]{grigoryan2019hsaCu}%
  \BibitemOpen
  \bibfield  {author} {\bibinfo {author} {\bibfnamefont {V.~G.}\ \bibnamefont
  {Grigoryan}}\ and\ \bibinfo {author} {\bibfnamefont {M.}~\bibnamefont
  {Springborg}},\ }\href {\doibase 10.1039/c9cp00123a} {\bibfield  {journal}
  {\bibinfo  {journal} {Phys. Chem. Chem. Phys.}\ }\textbf {\bibinfo {volume}
  {21}},\ \bibinfo {pages} {5646} (\bibinfo {year} {2019})}\BibitemShut
  {NoStop}%
\bibitem [{\citenamefont {Panizon}\ and\ \citenamefont
  {Ferrando}(2015)}]{panizon2015hsaPdPt}%
  \BibitemOpen
  \bibfield  {author} {\bibinfo {author} {\bibfnamefont {E.}~\bibnamefont
  {Panizon}}\ and\ \bibinfo {author} {\bibfnamefont {R.}~\bibnamefont
  {Ferrando}},\ }\href {\doibase 10.1103/PhysRevB.92.205417} {\bibfield
  {journal} {\bibinfo  {journal} {Phys. Rev. B}\ }\textbf {\bibinfo {volume}
  {92}},\ \bibinfo {pages} {205417} (\bibinfo {year} {2015})}\BibitemShut
  {NoStop}%
\bibitem [{\citenamefont {Bonventre}, \citenamefont {Panizon},\ and\
  \citenamefont {Ferrando}(2018)}]{bonventre2018hsaAgCuANi}%
  \BibitemOpen
  \bibfield  {author} {\bibinfo {author} {\bibfnamefont {D.}~\bibnamefont
  {Bonventre}}, \bibinfo {author} {\bibfnamefont {E.}~\bibnamefont {Panizon}},
  \ and\ \bibinfo {author} {\bibfnamefont {R.}~\bibnamefont {Ferrando}},\
  }\href {\doibase 10.1002/ppsc.201700425} {\bibfield  {journal} {\bibinfo
  {journal} {Part. Part. Syst. Charact.}\ }\textbf {\bibinfo {volume} {35}},\
  \bibinfo {pages} {1700425} (\bibinfo {year} {2018})}\BibitemShut {NoStop}%
\bibitem [{\citenamefont {Earl}\ and\ \citenamefont {Deem}(2005)}]{earl2005PT}%
  \BibitemOpen
  \bibfield  {author} {\bibinfo {author} {\bibfnamefont {D.~J.}\ \bibnamefont
  {Earl}}\ and\ \bibinfo {author} {\bibfnamefont {M.~W.}\ \bibnamefont
  {Deem}},\ }\href {\doibase 10.1039/b509983h} {\bibfield  {journal} {\bibinfo
  {journal} {Phys. Chem. Chem. Phys.}\ }\textbf {\bibinfo {volume} {7}},\
  \bibinfo {pages} {3910} (\bibinfo {year} {2005})}\BibitemShut {NoStop}%
\bibitem [{\citenamefont {Neirotti}\ \emph {et~al.}(2000)\citenamefont
  {Neirotti}, \citenamefont {Calvo}, \citenamefont {Freeman},\ and\
  \citenamefont {Doll}}]{neirotti2002ptmcLJ1}%
  \BibitemOpen
  \bibfield  {author} {\bibinfo {author} {\bibfnamefont {J.~P.}\ \bibnamefont
  {Neirotti}}, \bibinfo {author} {\bibfnamefont {F.}~\bibnamefont {Calvo}},
  \bibinfo {author} {\bibfnamefont {D.~L.}\ \bibnamefont {Freeman}}, \ and\
  \bibinfo {author} {\bibfnamefont {J.~D.}\ \bibnamefont {Doll}},\ }\href
  {\doibase 10.1063/1.481671} {\bibfield  {journal} {\bibinfo  {journal} {J.
  Chem. Phys.}\ }\textbf {\bibinfo {volume} {112}},\ \bibinfo {pages} {10340}
  (\bibinfo {year} {2000})}\BibitemShut {NoStop}%
\bibitem [{\citenamefont {Calvo}\ \emph {et~al.}(2000)\citenamefont {Calvo},
  \citenamefont {Neirotti}, \citenamefont {Freeman},\ and\ \citenamefont
  {Doll}}]{neirotti2002ptmcLJ2}%
  \BibitemOpen
  \bibfield  {author} {\bibinfo {author} {\bibfnamefont {F.}~\bibnamefont
  {Calvo}}, \bibinfo {author} {\bibfnamefont {J.~P.}\ \bibnamefont {Neirotti}},
  \bibinfo {author} {\bibfnamefont {D.~L.}\ \bibnamefont {Freeman}}, \ and\
  \bibinfo {author} {\bibfnamefont {J.~D.}\ \bibnamefont {Doll}},\ }\href
  {\doibase 10.1063/1.481672} {\bibfield  {journal} {\bibinfo  {journal} {J.
  Chem. Phys.}\ }\textbf {\bibinfo {volume} {112}},\ \bibinfo {pages} {10350}
  (\bibinfo {year} {2000})}\BibitemShut {NoStop}%
\bibitem [{\citenamefont {Ballard}\ and\ \citenamefont
  {Wales}(2014)}]{ballard2014}%
  \BibitemOpen
  \bibfield  {author} {\bibinfo {author} {\bibfnamefont {A.~J.}\ \bibnamefont
  {Ballard}}\ and\ \bibinfo {author} {\bibfnamefont {D.~J.}\ \bibnamefont
  {Wales}},\ }\href {\doibase 10.1021/ct500797a} {\bibfield  {journal}
  {\bibinfo  {journal} {J. Chem. Theory Comput.}\ }\textbf {\bibinfo {volume}
  {10}},\ \bibinfo {pages} {5599} (\bibinfo {year} {2014})}\BibitemShut
  {NoStop}%
\bibitem [{\citenamefont {Guimarães}\ \emph {et~al.}(2020)\citenamefont
  {Guimarães}, \citenamefont {de~Almeida}, \citenamefont {Marques},\ and\
  \citenamefont {Prudente}}]{guimarães2020ptmcSolvation}%
  \BibitemOpen
  \bibfield  {author} {\bibinfo {author} {\bibfnamefont {M.~N.}\ \bibnamefont
  {Guimarães}}, \bibinfo {author} {\bibfnamefont {M.~M.}\ \bibnamefont
  {de~Almeida}}, \bibinfo {author} {\bibfnamefont {J.~M.~C.}\ \bibnamefont
  {Marques}}, \ and\ \bibinfo {author} {\bibfnamefont {F.~V.}\ \bibnamefont
  {Prudente}},\ }\href {\doibase 10.1039/D0CP01283A} {\bibfield  {journal}
  {\bibinfo  {journal} {Phys. Chem. Chem. Phys.}\ }\textbf {\bibinfo {volume}
  {22}},\ \bibinfo {pages} {10882} (\bibinfo {year} {2020})}\BibitemShut
  {NoStop}%
\bibitem [{\citenamefont {Shu}\ \emph {et~al.}(2012)\citenamefont {Shu},
  \citenamefont {Yang}, \citenamefont {Zhai}, \citenamefont {Sun},
  \citenamefont {Xiang},\ and\ \citenamefont {Gong}}]{calvo2012PTMDnFeMelting}%
  \BibitemOpen
  \bibfield  {author} {\bibinfo {author} {\bibfnamefont {Q.}~\bibnamefont
  {Shu}}, \bibinfo {author} {\bibfnamefont {Y.}~\bibnamefont {Yang}}, \bibinfo
  {author} {\bibfnamefont {Y.}~\bibnamefont {Zhai}}, \bibinfo {author}
  {\bibfnamefont {D.~Y.}\ \bibnamefont {Sun}}, \bibinfo {author} {\bibfnamefont
  {H.~J.}\ \bibnamefont {Xiang}}, \ and\ \bibinfo {author} {\bibfnamefont
  {X.~G.}\ \bibnamefont {Gong}},\ }\href {\doibase 10.1039/c2nr30853c}
  {\bibfield  {journal} {\bibinfo  {journal} {Nanoscale}\ }\textbf {\bibinfo
  {volume} {4}},\ \bibinfo {pages} {6307} (\bibinfo {year} {2012})}\BibitemShut
  {NoStop}%
\bibitem [{\citenamefont {Tarrat}, \citenamefont {Rapacioli},\ and\
  \citenamefont {Spiegelman}(2018)}]{tarrat2018PTMDnAu}%
  \BibitemOpen
  \bibfield  {author} {\bibinfo {author} {\bibfnamefont {N.}~\bibnamefont
  {Tarrat}}, \bibinfo {author} {\bibfnamefont {M.}~\bibnamefont {Rapacioli}}, \
  and\ \bibinfo {author} {\bibfnamefont {F.}~\bibnamefont {Spiegelman}},\
  }\href {\doibase 10.1063/1.5021785} {\bibfield  {journal} {\bibinfo
  {journal} {J. Chem. Phys.}\ }\textbf {\bibinfo {volume} {148}},\ \bibinfo
  {pages} {204308} (\bibinfo {year} {2018})}\BibitemShut {NoStop}%
\bibitem [{\citenamefont {Nelli}, \citenamefont {Mottet},\ and\ \citenamefont
  {Ferrando}(2023)}]{nelli2023FarDiss}%
  \BibitemOpen
  \bibfield  {author} {\bibinfo {author} {\bibfnamefont {D.}~\bibnamefont
  {Nelli}}, \bibinfo {author} {\bibfnamefont {C.}~\bibnamefont {Mottet}}, \
  and\ \bibinfo {author} {\bibfnamefont {R.}~\bibnamefont {Ferrando}},\ }\href
  {\doibase 10.1039/D2FD00113F} {\bibfield  {journal} {\bibinfo  {journal}
  {Faraday Discuss.}\ }\textbf {\bibinfo {volume} {242}},\ \bibinfo {pages}
  {52} (\bibinfo {year} {2023})}\BibitemShut {NoStop}%
\bibitem [{\citenamefont {Nelli}\ \emph {et~al.}(2020)\citenamefont {Nelli},
  \citenamefont {Rossi}, \citenamefont {Wang}, \citenamefont {Palmer},\ and\
  \citenamefont {Ferrando}}]{nelli2020AuBHMC}%
  \BibitemOpen
  \bibfield  {author} {\bibinfo {author} {\bibfnamefont {D.}~\bibnamefont
  {Nelli}}, \bibinfo {author} {\bibfnamefont {G.}~\bibnamefont {Rossi}},
  \bibinfo {author} {\bibfnamefont {Z.}~\bibnamefont {Wang}}, \bibinfo {author}
  {\bibfnamefont {R.~E.}\ \bibnamefont {Palmer}}, \ and\ \bibinfo {author}
  {\bibfnamefont {R.}~\bibnamefont {Ferrando}},\ }\href {\doibase
  10.1039/c9nr10163b} {\bibfield  {journal} {\bibinfo  {journal} {Nanoscale}\
  }\textbf {\bibinfo {volume} {12}},\ \bibinfo {pages} {7688} (\bibinfo {year}
  {2020})}\BibitemShut {NoStop}%
\bibitem [{\citenamefont {Li}\ \emph {et~al.}(2015)\citenamefont {Li},
  \citenamefont {Li}, \citenamefont {Pedersen}, \citenamefont {Gao},
  \citenamefont {Khetrapal}, \citenamefont {Jonsson},\ and\ \citenamefont
  {Zeng}}]{li2015AuCatShapeEffect}%
  \BibitemOpen
  \bibfield  {author} {\bibinfo {author} {\bibfnamefont {H.}~\bibnamefont
  {Li}}, \bibinfo {author} {\bibfnamefont {L.}~\bibnamefont {Li}}, \bibinfo
  {author} {\bibfnamefont {A.}~\bibnamefont {Pedersen}}, \bibinfo {author}
  {\bibfnamefont {Y.}~\bibnamefont {Gao}}, \bibinfo {author} {\bibfnamefont
  {N.}~\bibnamefont {Khetrapal}}, \bibinfo {author} {\bibfnamefont
  {H.}~\bibnamefont {Jonsson}}, \ and\ \bibinfo {author} {\bibfnamefont
  {X.~C.}\ \bibnamefont {Zeng}},\ }\href {\doibase 10.1021/nl504192u}
  {\bibfield  {journal} {\bibinfo  {journal} {Nano Lett.}\ }\textbf {\bibinfo
  {volume} {15}},\ \bibinfo {pages} {682} (\bibinfo {year} {2015})}\BibitemShut
  {NoStop}%
\bibitem [{\citenamefont {Zhao}\ \emph {et~al.}(2017)\citenamefont {Zhao},
  \citenamefont {Zhang}, \citenamefont {Huang},\ and\ \citenamefont
  {Wang}}]{zhao2017sizeCatCu}%
  \BibitemOpen
  \bibfield  {author} {\bibinfo {author} {\bibfnamefont {B.}~\bibnamefont
  {Zhao}}, \bibinfo {author} {\bibfnamefont {R.}~\bibnamefont {Zhang}},
  \bibinfo {author} {\bibfnamefont {Z.}~\bibnamefont {Huang}}, \ and\ \bibinfo
  {author} {\bibfnamefont {B.}~\bibnamefont {Wang}},\ }\href {\doibase
  10.1016/j.apcata.2017.08.001} {\bibfield  {journal} {\bibinfo  {journal}
  {Appl. Catal. A: Gen.}\ }\textbf {\bibinfo {volume} {546}},\ \bibinfo {pages}
  {111} (\bibinfo {year} {2017})}\BibitemShut {NoStop}%
\bibitem [{\citenamefont {Jørgensen}\ and\ \citenamefont
  {Grönbeck}(2018)}]{jørgensen2018siteAssembly}%
  \BibitemOpen
  \bibfield  {author} {\bibinfo {author} {\bibfnamefont {M.}~\bibnamefont
  {Jørgensen}}\ and\ \bibinfo {author} {\bibfnamefont {H.}~\bibnamefont
  {Grönbeck}},\ }\href {\doibase 10.1002/anie.201802113} {\bibfield  {journal}
  {\bibinfo  {journal} {Angew. Chem. Int. Ed.}\ }\textbf {\bibinfo {volume}
  {57}},\ \bibinfo {pages} {5086} (\bibinfo {year} {2018})}\BibitemShut
  {NoStop}%
\bibitem [{\citenamefont {Rong}\ \emph {et~al.}(2021)\citenamefont {Rong},
  \citenamefont {Zou}, \citenamefont {Zang}, \citenamefont {Xi}, \citenamefont
  {Wei}, \citenamefont {Long}, \citenamefont {Hu}, \citenamefont {Ji},\ and\
  \citenamefont {Duan}}]{rong2021sizeDepCat}%
  \BibitemOpen
  \bibfield  {author} {\bibinfo {author} {\bibfnamefont {W.}~\bibnamefont
  {Rong}}, \bibinfo {author} {\bibfnamefont {H.}~\bibnamefont {Zou}}, \bibinfo
  {author} {\bibfnamefont {W.}~\bibnamefont {Zang}}, \bibinfo {author}
  {\bibfnamefont {S.}~\bibnamefont {Xi}}, \bibinfo {author} {\bibfnamefont
  {S.}~\bibnamefont {Wei}}, \bibinfo {author} {\bibfnamefont {B.}~\bibnamefont
  {Long}}, \bibinfo {author} {\bibfnamefont {J.}~\bibnamefont {Hu}}, \bibinfo
  {author} {\bibfnamefont {Y.}~\bibnamefont {Ji}}, \ and\ \bibinfo {author}
  {\bibfnamefont {L.}~\bibnamefont {Duan}},\ }\href {\doibase
  10.1002/anie.202011836} {\bibfield  {journal} {\bibinfo  {journal} {Angew.
  Chem. Int. Ed.}\ }\textbf {\bibinfo {volume} {60}},\ \bibinfo {pages} {466}
  (\bibinfo {year} {2021})}\BibitemShut {NoStop}%
\bibitem [{\citenamefont {Rossi}, \citenamefont {Asara},\ and\ \citenamefont
  {Baletto}(2019)}]{rossi2019npGenome}%
  \BibitemOpen
  \bibfield  {author} {\bibinfo {author} {\bibfnamefont {K.}~\bibnamefont
  {Rossi}}, \bibinfo {author} {\bibfnamefont {G.~G.}\ \bibnamefont {Asara}}, \
  and\ \bibinfo {author} {\bibfnamefont {F.}~\bibnamefont {Baletto}},\ }\href
  {\doibase 10.1039/C8CP05720F} {\bibfield  {journal} {\bibinfo  {journal}
  {Phys. Chem. Chem. Phys.}\ }\textbf {\bibinfo {volume} {21}},\ \bibinfo
  {pages} {4888} (\bibinfo {year} {2019})}\BibitemShut {NoStop}%
\bibitem [{\citenamefont {Cheula}, \citenamefont {Maestri},\ and\ \citenamefont
  {Mpourmpakis}(2020)}]{cheula2020npEnsembleCat}%
  \BibitemOpen
  \bibfield  {author} {\bibinfo {author} {\bibfnamefont {R.}~\bibnamefont
  {Cheula}}, \bibinfo {author} {\bibfnamefont {M.}~\bibnamefont {Maestri}}, \
  and\ \bibinfo {author} {\bibfnamefont {G.}~\bibnamefont {Mpourmpakis}},\
  }\href {\doibase 10.1021/acscatal.0c01005} {\bibfield  {journal} {\bibinfo
  {journal} {ACS Catal.}\ }\textbf {\bibinfo {volume} {10}},\ \bibinfo {pages}
  {6149} (\bibinfo {year} {2020})}\BibitemShut {NoStop}%
\bibitem [{\citenamefont {Daw}\ and\ \citenamefont
  {Baskes}(1984)}]{DawPRB1984}%
  \BibitemOpen
  \bibfield  {author} {\bibinfo {author} {\bibfnamefont {M.~S.}\ \bibnamefont
  {Daw}}\ and\ \bibinfo {author} {\bibfnamefont {M.~I.}\ \bibnamefont
  {Baskes}},\ }\href {\doibase 10.1103/PhysRevB.29.6443} {\bibfield  {journal}
  {\bibinfo  {journal} {Phys. Rev. B}\ }\textbf {\bibinfo {volume} {29}},\
  \bibinfo {pages} {6443} (\bibinfo {year} {1984})}\BibitemShut {NoStop}%
\bibitem [{\citenamefont {Gupta}(1981)}]{gupta1981}%
  \BibitemOpen
  \bibfield  {author} {\bibinfo {author} {\bibfnamefont {R.~P.}\ \bibnamefont
  {Gupta}},\ }\href {\doibase 10.1103/PhysRevB.23.6265} {\bibfield  {journal}
  {\bibinfo  {journal} {Phys. Rev. B}\ }\textbf {\bibinfo {volume} {23}},\
  \bibinfo {pages} {62} (\bibinfo {year} {1981})}\BibitemShut {NoStop}%
\bibitem [{\citenamefont {Sutton}\ and\ \citenamefont
  {J.Chen}(1990)}]{suttonChen1990}%
  \BibitemOpen
  \bibfield  {author} {\bibinfo {author} {\bibfnamefont {A.~P.}\ \bibnamefont
  {Sutton}}\ and\ \bibinfo {author} {\bibnamefont {J.Chen}},\ }\href {\doibase
  10.1080/09500839008206493} {\bibfield  {journal} {\bibinfo  {journal}
  {Philos. Mag. Lett.}\ }\textbf {\bibinfo {volume} {61}},\ \bibinfo {pages}
  {139} (\bibinfo {year} {1990})}\BibitemShut {NoStop}%
\bibitem [{\citenamefont {Pyykko}(2004)}]{pyykko2004AuRelEffects}%
  \BibitemOpen
  \bibfield  {author} {\bibinfo {author} {\bibfnamefont {P.}~\bibnamefont
  {Pyykko}},\ }\href {\doibase 10.1002/anie.200300624} {\bibfield  {journal}
  {\bibinfo  {journal} {Angew. Chem. Int. Ed.}\ }\textbf {\bibinfo {volume}
  {43}},\ \bibinfo {pages} {4412} (\bibinfo {year} {2004})}\BibitemShut
  {NoStop}%
\bibitem [{\citenamefont {Furche}\ \emph {et~al.}(2002)\citenamefont {Furche},
  \citenamefont {Ahlrichs}, \citenamefont {Weis}, \citenamefont {Jacob},
  \citenamefont {Gilb}, \citenamefont {Bierweiler},\ and\ \citenamefont
  {Kappes}}]{furche2002AuPlanarStructsExpIon}%
  \BibitemOpen
  \bibfield  {author} {\bibinfo {author} {\bibfnamefont {F.}~\bibnamefont
  {Furche}}, \bibinfo {author} {\bibfnamefont {R.}~\bibnamefont {Ahlrichs}},
  \bibinfo {author} {\bibfnamefont {P.}~\bibnamefont {Weis}}, \bibinfo {author}
  {\bibfnamefont {C.}~\bibnamefont {Jacob}}, \bibinfo {author} {\bibfnamefont
  {S.}~\bibnamefont {Gilb}}, \bibinfo {author} {\bibfnamefont {T.}~\bibnamefont
  {Bierweiler}}, \ and\ \bibinfo {author} {\bibfnamefont {M.~M.}\ \bibnamefont
  {Kappes}},\ }\href {\doibase 10.1063/1.1507582} {\bibfield  {journal}
  {\bibinfo  {journal} {J. Chem. Phys.}\ }\textbf {\bibinfo {volume} {117}},\
  \bibinfo {pages} {6982} (\bibinfo {year} {2002})}\BibitemShut {NoStop}%
\bibitem [{\citenamefont {Hakkinen}, \citenamefont {Moseler},\ and\
  \citenamefont {Landman}(2002)}]{hakkinen2002Au2Dto3DCuAg}%
  \BibitemOpen
  \bibfield  {author} {\bibinfo {author} {\bibfnamefont {H.}~\bibnamefont
  {Hakkinen}}, \bibinfo {author} {\bibfnamefont {M.}~\bibnamefont {Moseler}}, \
  and\ \bibinfo {author} {\bibfnamefont {U.}~\bibnamefont {Landman}},\ }\href
  {\doibase 10.1103/PhysRevLett.89.033401} {\bibfield  {journal} {\bibinfo
  {journal} {Phys. Rev. Lett.}\ }\textbf {\bibinfo {volume} {89}},\ \bibinfo
  {pages} {033401} (\bibinfo {year} {2002})}\BibitemShut {NoStop}%
\bibitem [{\citenamefont {Hakkinen}\ \emph {et~al.}(2003)\citenamefont
  {Hakkinen}, \citenamefont {Yoon}, \citenamefont {Landman}, \citenamefont
  {Li}, \citenamefont {Zhai},\ and\ \citenamefont
  {Wang}}]{hakkinen2003AuPlanarStructs}%
  \BibitemOpen
  \bibfield  {author} {\bibinfo {author} {\bibfnamefont {H.}~\bibnamefont
  {Hakkinen}}, \bibinfo {author} {\bibfnamefont {B.}~\bibnamefont {Yoon}},
  \bibinfo {author} {\bibfnamefont {U.}~\bibnamefont {Landman}}, \bibinfo
  {author} {\bibfnamefont {X.}~\bibnamefont {Li}}, \bibinfo {author}
  {\bibfnamefont {H.}~\bibnamefont {Zhai}}, \ and\ \bibinfo {author}
  {\bibfnamefont {L.}~\bibnamefont {Wang}},\ }\href {\doibase
  10.1021/jp035437i} {\bibfield  {journal} {\bibinfo  {journal} {J. Phys. Chem.
  A}\ }\textbf {\bibinfo {volume} {107}},\ \bibinfo {pages} {6168} (\bibinfo
  {year} {2003})}\BibitemShut {NoStop}%
\bibitem [{\citenamefont {Johansson}, \citenamefont {Sundholm},\ and\
  \citenamefont {Vaara}(2004)}]{johansson2004AuCageStructs}%
  \BibitemOpen
  \bibfield  {author} {\bibinfo {author} {\bibfnamefont {M.~P.}\ \bibnamefont
  {Johansson}}, \bibinfo {author} {\bibfnamefont {D.}~\bibnamefont {Sundholm}},
  \ and\ \bibinfo {author} {\bibfnamefont {J.}~\bibnamefont {Vaara}},\ }\href
  {\doibase 10.1002/anie.200453986} {\bibfield  {journal} {\bibinfo  {journal}
  {Angew. Chem. Int. Ed.}\ }\textbf {\bibinfo {volume} {43}},\ \bibinfo {pages}
  {2678} (\bibinfo {year} {2004})}\BibitemShut {NoStop}%
\bibitem [{\citenamefont {Gu}\ \emph {et~al.}(2004)\citenamefont {Gu},
  \citenamefont {Ji}, \citenamefont {Wei},\ and\ \citenamefont
  {Gong}}]{gu2004AuCageStructs}%
  \BibitemOpen
  \bibfield  {author} {\bibinfo {author} {\bibfnamefont {X.}~\bibnamefont
  {Gu}}, \bibinfo {author} {\bibfnamefont {M.}~\bibnamefont {Ji}}, \bibinfo
  {author} {\bibfnamefont {S.~H.}\ \bibnamefont {Wei}}, \ and\ \bibinfo
  {author} {\bibfnamefont {X.~G.}\ \bibnamefont {Gong}},\ }\href {\doibase
  10.1103/PhysRevB.70.205401} {\bibfield  {journal} {\bibinfo  {journal} {Phys.
  Rev. B}\ }\textbf {\bibinfo {volume} {70}},\ \bibinfo {pages} {205401}
  (\bibinfo {year} {2004})}\BibitemShut {NoStop}%
\bibitem [{\citenamefont {Fa}\ and\ \citenamefont
  {Dong}(2006)}]{fa2006AuCageStructs}%
  \BibitemOpen
  \bibfield  {author} {\bibinfo {author} {\bibfnamefont {W.}~\bibnamefont
  {Fa}}\ and\ \bibinfo {author} {\bibfnamefont {J.}~\bibnamefont {Dong}},\
  }\href {\doibase 10.1063/1.2179071} {\bibfield  {journal} {\bibinfo
  {journal} {J. Chem. Phys.}\ }\textbf {\bibinfo {volume} {124}},\ \bibinfo
  {pages} {114310} (\bibinfo {year} {2006})}\BibitemShut {NoStop}%
\bibitem [{\citenamefont {Xing}\ \emph {et~al.}(2006)\citenamefont {Xing},
  \citenamefont {Yoon}, \citenamefont {Landman},\ and\ \citenamefont
  {Parks}}]{xing2006AuCageStructs}%
  \BibitemOpen
  \bibfield  {author} {\bibinfo {author} {\bibfnamefont {X.}~\bibnamefont
  {Xing}}, \bibinfo {author} {\bibfnamefont {B.}~\bibnamefont {Yoon}}, \bibinfo
  {author} {\bibfnamefont {U.}~\bibnamefont {Landman}}, \ and\ \bibinfo
  {author} {\bibfnamefont {J.~H.}\ \bibnamefont {Parks}},\ }\href {\doibase
  10.1103/PhysRevB.74.165423} {\bibfield  {journal} {\bibinfo  {journal} {Phys.
  Rev. B}\ }\textbf {\bibinfo {volume} {74}},\ \bibinfo {pages} {165423}
  (\bibinfo {year} {2006})}\BibitemShut {NoStop}%
\bibitem [{\citenamefont {Mottet}\ \emph {et~al.}(2005)\citenamefont {Mottet},
  \citenamefont {Rossi}, \citenamefont {Baletto},\ and\ \citenamefont
  {Ferrando}}]{mottet2005rosette}%
  \BibitemOpen
  \bibfield  {author} {\bibinfo {author} {\bibfnamefont {C.}~\bibnamefont
  {Mottet}}, \bibinfo {author} {\bibfnamefont {G.}~\bibnamefont {Rossi}},
  \bibinfo {author} {\bibfnamefont {F.}~\bibnamefont {Baletto}}, \ and\
  \bibinfo {author} {\bibfnamefont {R.}~\bibnamefont {Ferrando}},\ }\href
  {\doibase 10.1103/PhysRevLett.95.035501} {\bibfield  {journal} {\bibinfo
  {journal} {Phys. Rev. Lett.}\ }\textbf {\bibinfo {volume} {95}},\ \bibinfo
  {pages} {035501} (\bibinfo {year} {2005})}\BibitemShut {NoStop}%
\bibitem [{\citenamefont {Rossi}\ \emph {et~al.}(2018)\citenamefont {Rossi},
  \citenamefont {Pavan}, \citenamefont {Soon},\ and\ \citenamefont
  {Baletto}}]{rossi2018structTrans}%
  \BibitemOpen
  \bibfield  {author} {\bibinfo {author} {\bibfnamefont {K.}~\bibnamefont
  {Rossi}}, \bibinfo {author} {\bibfnamefont {L.}~\bibnamefont {Pavan}},
  \bibinfo {author} {\bibfnamefont {Y.~Y.}\ \bibnamefont {Soon}}, \ and\
  \bibinfo {author} {\bibfnamefont {F.}~\bibnamefont {Baletto}},\ }\href
  {\doibase 10.1140/epjb/e2017-80281-6} {\bibfield  {journal} {\bibinfo
  {journal} {Eur. Phys. J. B}\ }\textbf {\bibinfo {volume} {91}},\ \bibinfo
  {pages} {33} (\bibinfo {year} {2018})}\BibitemShut {NoStop}%
\bibitem [{\citenamefont {Apra}\ \emph {et~al.}(2004)\citenamefont {Apra},
  \citenamefont {Baletto}, \citenamefont {Ferrando},\ and\ \citenamefont
  {Fortunelli}}]{apra2004AuRosette}%
  \BibitemOpen
  \bibfield  {author} {\bibinfo {author} {\bibfnamefont {E.}~\bibnamefont
  {Apra}}, \bibinfo {author} {\bibfnamefont {F.}~\bibnamefont {Baletto}},
  \bibinfo {author} {\bibfnamefont {R.}~\bibnamefont {Ferrando}}, \ and\
  \bibinfo {author} {\bibfnamefont {A.}~\bibnamefont {Fortunelli}},\ }\href
  {\doibase 10.1103/PhysRevLett.93.065502} {\bibfield  {journal} {\bibinfo
  {journal} {Phys. Rev. Lett.}\ }\textbf {\bibinfo {volume} {93}},\ \bibinfo
  {pages} {065502} (\bibinfo {year} {2004})}\BibitemShut {NoStop}%
\bibitem [{\citenamefont {{D. Nelli}}(2022)}]{Nelli2022epjap}%
  \BibitemOpen
  \bibfield  {author} {\bibinfo {author} {\bibnamefont {{D. Nelli}}},\ }\href
  {\doibase 10.1051/epjap/2022210282} {\bibfield  {journal} {\bibinfo
  {journal} {Eur. Phys. J. Appl. Phys.}\ }\textbf {\bibinfo {volume} {97}},\
  \bibinfo {pages} {18} (\bibinfo {year} {2022})}\BibitemShut {NoStop}%
\bibitem [{\citenamefont {Yin}\ and\ \citenamefont
  {Luo}(2021)}]{yin2021coinageRev}%
  \BibitemOpen
  \bibfield  {author} {\bibinfo {author} {\bibfnamefont {B.}~\bibnamefont
  {Yin}}\ and\ \bibinfo {author} {\bibfnamefont {Z.}~\bibnamefont {Luo}},\
  }\href {\doibase 10.1016/j.ccr.2020.213643} {\bibfield  {journal} {\bibinfo
  {journal} {Coord. Chem. Rev.}\ }\textbf {\bibinfo {volume} {429}},\ \bibinfo
  {pages} {213643} (\bibinfo {year} {2021})}\BibitemShut {NoStop}%
\bibitem [{\citenamefont {Roldán}\ \emph {et~al.}(2008)\citenamefont
  {Roldán}, \citenamefont {Viñes}, \citenamefont {Illas}, \citenamefont
  {Ricart},\ and\ \citenamefont {Neyman}}]{roldan2008DFTfcc}%
  \BibitemOpen
  \bibfield  {author} {\bibinfo {author} {\bibfnamefont {A.}~\bibnamefont
  {Roldán}}, \bibinfo {author} {\bibfnamefont {F.}~\bibnamefont {Viñes}},
  \bibinfo {author} {\bibfnamefont {F.}~\bibnamefont {Illas}}, \bibinfo
  {author} {\bibfnamefont {J.~M.}\ \bibnamefont {Ricart}}, \ and\ \bibinfo
  {author} {\bibfnamefont {K.~M.}\ \bibnamefont {Neyman}},\ }\href {\doibase
  10.1007/s00214-008-0423-x} {\bibfield  {journal} {\bibinfo  {journal} {Theor.
  Chem. Acc.}\ }\textbf {\bibinfo {volume} {120}},\ \bibinfo {pages} {565}
  (\bibinfo {year} {2008})}\BibitemShut {NoStop}%
\bibitem [{\citenamefont {Kiss}, \citenamefont {Miotto},\ and\ \citenamefont
  {Ferraz}(2011)}]{kiss2011AgDFTSizeEffects}%
  \BibitemOpen
  \bibfield  {author} {\bibinfo {author} {\bibfnamefont {F.~D.}\ \bibnamefont
  {Kiss}}, \bibinfo {author} {\bibfnamefont {R.}~\bibnamefont {Miotto}}, \ and\
  \bibinfo {author} {\bibfnamefont {A.~C.}\ \bibnamefont {Ferraz}},\ }\href
  {\doibase 10.1088/0957-4484/22/27/275708} {\bibfield  {journal} {\bibinfo
  {journal} {Nanotechnology}\ }\textbf {\bibinfo {volume} {22}},\ \bibinfo
  {pages} {275708} (\bibinfo {year} {2011})}\BibitemShut {NoStop}%
\bibitem [{\citenamefont {Oliveira}\ \emph {et~al.}(2016)\citenamefont
  {Oliveira}, \citenamefont {Tarrat}, \citenamefont {Cuny}, \citenamefont
  {Morillo},\ and\ \citenamefont {Lemoine}}]{oliveira2016benchmarkingDFTB}%
  \BibitemOpen
  \bibfield  {author} {\bibinfo {author} {\bibfnamefont {L.~F.~L.}\
  \bibnamefont {Oliveira}}, \bibinfo {author} {\bibfnamefont {N.}~\bibnamefont
  {Tarrat}}, \bibinfo {author} {\bibfnamefont {J.}~\bibnamefont {Cuny}},
  \bibinfo {author} {\bibfnamefont {J.}~\bibnamefont {Morillo}}, \ and\
  \bibinfo {author} {\bibfnamefont {D.}~\bibnamefont {Lemoine}},\ }\href
  {\doibase 10.1021/acs.jpca.6b09292} {\bibfield  {journal} {\bibinfo
  {journal} {J. Phys. Chem. A}\ }\textbf {\bibinfo {volume} {120}},\ \bibinfo
  {pages} {8469} (\bibinfo {year} {2016})}\BibitemShut {NoStop}%
\bibitem [{\citenamefont {Langlois}\ \emph {et~al.}(2008)\citenamefont
  {Langlois}, \citenamefont {Alloyeau}, \citenamefont {Bouar}, \citenamefont
  {Loiseau}, \citenamefont {Oikawa}, \citenamefont {Mottet},\ and\
  \citenamefont {Ricolleau}}]{langlois2008CuHRTEM}%
  \BibitemOpen
  \bibfield  {author} {\bibinfo {author} {\bibfnamefont {C.}~\bibnamefont
  {Langlois}}, \bibinfo {author} {\bibfnamefont {D.}~\bibnamefont {Alloyeau}},
  \bibinfo {author} {\bibfnamefont {Y.}~\bibnamefont {Bouar}}, \bibinfo
  {author} {\bibfnamefont {A.}~\bibnamefont {Loiseau}}, \bibinfo {author}
  {\bibfnamefont {T.}~\bibnamefont {Oikawa}}, \bibinfo {author} {\bibfnamefont
  {C.}~\bibnamefont {Mottet}}, \ and\ \bibinfo {author} {\bibfnamefont
  {C.}~\bibnamefont {Ricolleau}},\ }\href {\doibase 10.1039/B705912B}
  {\bibfield  {journal} {\bibinfo  {journal} {Faraday Discuss.}\ }\textbf
  {\bibinfo {volume} {138}},\ \bibinfo {pages} {375} (\bibinfo {year}
  {2008})}\BibitemShut {NoStop}%
\bibitem [{\citenamefont {Volk}\ \emph {et~al.}(2013)\citenamefont {Volk},
  \citenamefont {Thaler}, \citenamefont {Koch}, \citenamefont {Fisslthaler},
  \citenamefont {Grogger},\ and\ \citenamefont {Ernst}}]{volk2013AgTEM}%
  \BibitemOpen
  \bibfield  {author} {\bibinfo {author} {\bibfnamefont {A.}~\bibnamefont
  {Volk}}, \bibinfo {author} {\bibfnamefont {P.}~\bibnamefont {Thaler}},
  \bibinfo {author} {\bibfnamefont {M.}~\bibnamefont {Koch}}, \bibinfo {author}
  {\bibfnamefont {E.}~\bibnamefont {Fisslthaler}}, \bibinfo {author}
  {\bibfnamefont {W.}~\bibnamefont {Grogger}}, \ and\ \bibinfo {author}
  {\bibfnamefont {W.~E.}\ \bibnamefont {Ernst}},\ }\href {\doibase
  10.1063/1.4807843} {\bibfield  {journal} {\bibinfo  {journal} {J. Chem.
  Phys.}\ }\textbf {\bibinfo {volume} {138}},\ \bibinfo {pages} {214312}
  (\bibinfo {year} {2013})}\BibitemShut {NoStop}%
\bibitem [{\citenamefont {Baletto}, \citenamefont {Mottet},\ and\ \citenamefont
  {Ferrando}(2001)}]{Baletto2001prb}%
  \BibitemOpen
  \bibfield  {author} {\bibinfo {author} {\bibfnamefont {F.}~\bibnamefont
  {Baletto}}, \bibinfo {author} {\bibfnamefont {C.}~\bibnamefont {Mottet}}, \
  and\ \bibinfo {author} {\bibfnamefont {R.}~\bibnamefont {Ferrando}},\
  }\href@noop {} {\bibfield  {journal} {\bibinfo  {journal} {Phys. Rev. B}\
  }\textbf {\bibinfo {volume} {63}},\ \bibinfo {pages} {155408} (\bibinfo
  {year} {2001})}\BibitemShut {NoStop}%
\bibitem [{\citenamefont {El~koraychy}\ \emph {et~al.}(2022)\citenamefont
  {El~koraychy}, \citenamefont {Roncaglia}, \citenamefont {Nelli},
  \citenamefont {Cerbelaud},\ and\ \citenamefont
  {Ferrando}}]{Elkoraychy2022nh}%
  \BibitemOpen
  \bibfield  {author} {\bibinfo {author} {\bibfnamefont {E.~y.}\ \bibnamefont
  {El~koraychy}}, \bibinfo {author} {\bibfnamefont {C.}~\bibnamefont
  {Roncaglia}}, \bibinfo {author} {\bibfnamefont {D.}~\bibnamefont {Nelli}},
  \bibinfo {author} {\bibfnamefont {M.}~\bibnamefont {Cerbelaud}}, \ and\
  \bibinfo {author} {\bibfnamefont {R.}~\bibnamefont {Ferrando}},\ }\href
  {\doibase 10.1039/D1NH00599E} {\bibfield  {journal} {\bibinfo  {journal}
  {Nanoscale Horiz.}\ }\textbf {\bibinfo {volume} {7}},\ \bibinfo {pages} {883}
  (\bibinfo {year} {2022})}\BibitemShut {NoStop}%
\bibitem [{\citenamefont {Loffreda}\ \emph {et~al.}(2021)\citenamefont
  {Loffreda}, \citenamefont {Foster}, \citenamefont {Palmer},\ and\
  \citenamefont {Tarrat}}]{loffreda2021Ag309TEM}%
  \BibitemOpen
  \bibfield  {author} {\bibinfo {author} {\bibfnamefont {D.}~\bibnamefont
  {Loffreda}}, \bibinfo {author} {\bibfnamefont {D.~M.}\ \bibnamefont
  {Foster}}, \bibinfo {author} {\bibfnamefont {R.~E.}\ \bibnamefont {Palmer}},
  \ and\ \bibinfo {author} {\bibfnamefont {N.}~\bibnamefont {Tarrat}},\ }\href
  {\doibase 10.1021/acs.jpclett.1c00259} {\bibfield  {journal} {\bibinfo
  {journal} {J. Phys. Chem. Lett.}\ }\textbf {\bibinfo {volume} {12}},\
  \bibinfo {pages} {3705} (\bibinfo {year} {2021})}\BibitemShut {NoStop}%
\bibitem [{\citenamefont {Wells}\ \emph {et~al.}(2015)\citenamefont {Wells},
  \citenamefont {Rossi}, \citenamefont {Ferrando},\ and\ \citenamefont
  {Palmer}}]{wells2015AuImagingACFraction}%
  \BibitemOpen
  \bibfield  {author} {\bibinfo {author} {\bibfnamefont {D.~M.}\ \bibnamefont
  {Wells}}, \bibinfo {author} {\bibfnamefont {G.}~\bibnamefont {Rossi}},
  \bibinfo {author} {\bibfnamefont {R.}~\bibnamefont {Ferrando}}, \ and\
  \bibinfo {author} {\bibfnamefont {R.~E.}\ \bibnamefont {Palmer}},\ }\href
  {\doibase 10.1039/c4nr05811a} {\bibfield  {journal} {\bibinfo  {journal}
  {Nanoscale}\ }\textbf {\bibinfo {volume} {7}},\ \bibinfo {pages} {6498}
  (\bibinfo {year} {2015})}\BibitemShut {NoStop}%
\bibitem [{\citenamefont {Foster}, \citenamefont {Ferrando},\ and\
  \citenamefont {Palmer}(2018)}]{foster2018AuImagingACFraction}%
  \BibitemOpen
  \bibfield  {author} {\bibinfo {author} {\bibfnamefont {D.~M.}\ \bibnamefont
  {Foster}}, \bibinfo {author} {\bibfnamefont {R.}~\bibnamefont {Ferrando}}, \
  and\ \bibinfo {author} {\bibfnamefont {R.~E.}\ \bibnamefont {Palmer}},\
  }\href {\doibase 10.1038/s41467-018-03794-9} {\bibfield  {journal} {\bibinfo
  {journal} {Nat. Commun.}\ }\textbf {\bibinfo {volume} {9}},\ \bibinfo {pages}
  {1323} (\bibinfo {year} {2018})}\BibitemShut {NoStop}%
\bibitem [{\citenamefont {Palomares-Baez}, \citenamefont {Panizon},\ and\
  \citenamefont {Ferrando}(2017)}]{palomaresBaez2017AuDFT}%
  \BibitemOpen
  \bibfield  {author} {\bibinfo {author} {\bibfnamefont {J.}~\bibnamefont
  {Palomares-Baez}}, \bibinfo {author} {\bibfnamefont {E.}~\bibnamefont
  {Panizon}}, \ and\ \bibinfo {author} {\bibfnamefont {R.}~\bibnamefont
  {Ferrando}},\ }\href {\doibase 10.1021/acs.nanolett.7b01994} {\bibfield
  {journal} {\bibinfo  {journal} {Nano Lett.}\ }\textbf {\bibinfo {volume}
  {17}},\ \bibinfo {pages} {5394} (\bibinfo {year} {2017})}\BibitemShut
  {NoStop}%
\bibitem [{\citenamefont {Cyrot-Lackmann}\ and\ \citenamefont
  {Ducastelle}(1971)}]{tbsma1971}%
  \BibitemOpen
  \bibfield  {author} {\bibinfo {author} {\bibfnamefont {F.}~\bibnamefont
  {Cyrot-Lackmann}}\ and\ \bibinfo {author} {\bibfnamefont {F.}~\bibnamefont
  {Ducastelle}},\ }\href {\doibase 10.1103/PhysRevB.4.2406} {\bibfield
  {journal} {\bibinfo  {journal} {Phys. Rev. B: Condens. Matter Mater. Phys.}\
  }\textbf {\bibinfo {volume} {4}},\ \bibinfo {pages} {2406} (\bibinfo {year}
  {1971})}\BibitemShut {NoStop}%
\bibitem [{\citenamefont {Rosato}, \citenamefont {Guillope},\ and\
  \citenamefont {Legrand}(1989)}]{rgl1989}%
  \BibitemOpen
  \bibfield  {author} {\bibinfo {author} {\bibfnamefont {V.}~\bibnamefont
  {Rosato}}, \bibinfo {author} {\bibfnamefont {M.}~\bibnamefont {Guillope}}, \
  and\ \bibinfo {author} {\bibfnamefont {B.}~\bibnamefont {Legrand}},\ }\href
  {\doibase 10.1080/01418618908205062} {\bibfield  {journal} {\bibinfo
  {journal} {Philos. Mag. A}\ }\textbf {\bibinfo {volume} {59}},\ \bibinfo
  {pages} {321} (\bibinfo {year} {1989})}\BibitemShut {NoStop}%
\bibitem [{\citenamefont {Han}, \citenamefont {Ferrando},\ and\ \citenamefont
  {Li}(2014)}]{han2014imagingOnMgO}%
  \BibitemOpen
  \bibfield  {author} {\bibinfo {author} {\bibfnamefont {Y.}~\bibnamefont
  {Han}}, \bibinfo {author} {\bibfnamefont {R.}~\bibnamefont {Ferrando}}, \
  and\ \bibinfo {author} {\bibfnamefont {Z.~Y.}\ \bibnamefont {Li}},\ }\href
  {\doibase 10.1021/jz4022975} {\bibfield  {journal} {\bibinfo  {journal} {J.
  Phys. Chem. Lett.}\ }\textbf {\bibinfo {volume} {5}},\ \bibinfo {pages} {131}
  (\bibinfo {year} {2014})}\BibitemShut {NoStop}%
\bibitem [{\citenamefont {Bochicchio}\ and\ \citenamefont
  {Ferrando}(2010)}]{BochicchioChiral2010}%
  \BibitemOpen
  \bibfield  {author} {\bibinfo {author} {\bibfnamefont {D.}~\bibnamefont
  {Bochicchio}}\ and\ \bibinfo {author} {\bibfnamefont {R.}~\bibnamefont
  {Ferrando}},\ }\href {\doibase 10.1021/nl102588p} {\bibfield  {journal}
  {\bibinfo  {journal} {Nano Lett.}\ }\textbf {\bibinfo {volume} {10}},\
  \bibinfo {pages} {4211} (\bibinfo {year} {2010})}\BibitemShut {NoStop}%
\bibitem [{\citenamefont {Panizon}\ \emph {et~al.}(2014)\citenamefont
  {Panizon}, \citenamefont {Bochicchio}, \citenamefont {Rossi},\ and\
  \citenamefont {Ferrando}}]{panizon2014diluteImpurity}%
  \BibitemOpen
  \bibfield  {author} {\bibinfo {author} {\bibfnamefont {E.}~\bibnamefont
  {Panizon}}, \bibinfo {author} {\bibfnamefont {D.}~\bibnamefont {Bochicchio}},
  \bibinfo {author} {\bibfnamefont {G.}~\bibnamefont {Rossi}}, \ and\ \bibinfo
  {author} {\bibfnamefont {R.}~\bibnamefont {Ferrando}},\ }\href {\doibase
  10.1021/cm501001f} {\bibfield  {journal} {\bibinfo  {journal} {Chem. Mater.}\
  }\textbf {\bibinfo {volume} {26}},\ \bibinfo {pages} {3354} (\bibinfo {year}
  {2014})}\BibitemShut {NoStop}%
\bibitem [{\citenamefont {Giannozzia}\ \emph {et~al.}(2009)\citenamefont
  {Giannozzia}, \citenamefont {Baroni}, \citenamefont {Bonini}, \citenamefont
  {Calandra}, \citenamefont {Car}, \citenamefont {Cavazzoni}, \citenamefont
  {Ceresoli}, \citenamefont {Chiarotti}, \citenamefont {Cococcioni},
  \citenamefont {Dabo}, \citenamefont {Corso}, \citenamefont {de~Gironcoli},
  \citenamefont {Fabris}, \citenamefont {Fratesi}, \citenamefont {Gebauer},
  \citenamefont {Gerstmann}, \citenamefont {Gougoussis}, \citenamefont
  {Kokalj}, \citenamefont {Lazzeri}, \citenamefont {Martin-Samos},
  \citenamefont {Marzari}, \citenamefont {Mauri}, \citenamefont {Mazzarello},
  \citenamefont {Paolini}, \citenamefont {Pasquarello}, \citenamefont
  {Paulatto}, \citenamefont {Sbraccia}, \citenamefont {Scandolo}, \citenamefont
  {Sclauzero}, \citenamefont {Seitsonen}, \citenamefont {Smogunov},
  \citenamefont {Umari},\ and\ \citenamefont {Wentzcovitch}}]{quantumEspresso}%
  \BibitemOpen
  \bibfield  {author} {\bibinfo {author} {\bibfnamefont {P.}~\bibnamefont
  {Giannozzia}}, \bibinfo {author} {\bibfnamefont {S.}~\bibnamefont {Baroni}},
  \bibinfo {author} {\bibfnamefont {N.}~\bibnamefont {Bonini}}, \bibinfo
  {author} {\bibfnamefont {M.}~\bibnamefont {Calandra}}, \bibinfo {author}
  {\bibfnamefont {R.}~\bibnamefont {Car}}, \bibinfo {author} {\bibfnamefont
  {C.}~\bibnamefont {Cavazzoni}}, \bibinfo {author} {\bibfnamefont
  {D.}~\bibnamefont {Ceresoli}}, \bibinfo {author} {\bibfnamefont {G.~L.}\
  \bibnamefont {Chiarotti}}, \bibinfo {author} {\bibfnamefont {M.}~\bibnamefont
  {Cococcioni}}, \bibinfo {author} {\bibfnamefont {I.}~\bibnamefont {Dabo}},
  \bibinfo {author} {\bibfnamefont {A.~D.}\ \bibnamefont {Corso}}, \bibinfo
  {author} {\bibfnamefont {S.}~\bibnamefont {de~Gironcoli}}, \bibinfo {author}
  {\bibfnamefont {S.}~\bibnamefont {Fabris}}, \bibinfo {author} {\bibfnamefont
  {G.}~\bibnamefont {Fratesi}}, \bibinfo {author} {\bibfnamefont
  {R.}~\bibnamefont {Gebauer}}, \bibinfo {author} {\bibfnamefont
  {U.}~\bibnamefont {Gerstmann}}, \bibinfo {author} {\bibfnamefont
  {C.}~\bibnamefont {Gougoussis}}, \bibinfo {author} {\bibfnamefont
  {A.}~\bibnamefont {Kokalj}}, \bibinfo {author} {\bibfnamefont
  {M.}~\bibnamefont {Lazzeri}}, \bibinfo {author} {\bibfnamefont
  {L.}~\bibnamefont {Martin-Samos}}, \bibinfo {author} {\bibfnamefont
  {N.}~\bibnamefont {Marzari}}, \bibinfo {author} {\bibfnamefont
  {F.}~\bibnamefont {Mauri}}, \bibinfo {author} {\bibfnamefont
  {R.}~\bibnamefont {Mazzarello}}, \bibinfo {author} {\bibfnamefont
  {S.}~\bibnamefont {Paolini}}, \bibinfo {author} {\bibfnamefont
  {A.}~\bibnamefont {Pasquarello}}, \bibinfo {author} {\bibfnamefont
  {L.}~\bibnamefont {Paulatto}}, \bibinfo {author} {\bibfnamefont
  {C.}~\bibnamefont {Sbraccia}}, \bibinfo {author} {\bibfnamefont
  {S.}~\bibnamefont {Scandolo}}, \bibinfo {author} {\bibfnamefont
  {G.}~\bibnamefont {Sclauzero}}, \bibinfo {author} {\bibfnamefont {A.~P.}\
  \bibnamefont {Seitsonen}}, \bibinfo {author} {\bibfnamefont {A.}~\bibnamefont
  {Smogunov}}, \bibinfo {author} {\bibfnamefont {P.}~\bibnamefont {Umari}}, \
  and\ \bibinfo {author} {\bibfnamefont {R.~M.}\ \bibnamefont {Wentzcovitch}},\
  }\href {\doibase 10.1088/0953-8984/21/39/395502} {\bibfield  {journal}
  {\bibinfo  {journal} {J. Phys.: Condens. Matter}\ }\textbf {\bibinfo {volume}
  {21}},\ \bibinfo {pages} {395502} (\bibinfo {year} {2009})}\BibitemShut
  {NoStop}%
\bibitem [{\citenamefont {Kresse}\ and\ \citenamefont
  {Furthmüller}(1996)}]{kresse1996VASPcms}%
  \BibitemOpen
  \bibfield  {author} {\bibinfo {author} {\bibfnamefont {G.}~\bibnamefont
  {Kresse}}\ and\ \bibinfo {author} {\bibfnamefont {J.}~\bibnamefont
  {Furthmüller}},\ }\href {\doibase 10.1016/0927-0256(96)00008-0} {\bibfield
  {journal} {\bibinfo  {journal} {Comput. Mat. Sci.}\ }\textbf {\bibinfo
  {volume} {6}},\ \bibinfo {pages} {15} (\bibinfo {year} {1996})}\BibitemShut
  {NoStop}%
\bibitem [{\citenamefont {Perdew}, \citenamefont {Burke},\ and\ \citenamefont
  {Ernzerhof}(1996)}]{pbeXCFunctional}%
  \BibitemOpen
  \bibfield  {author} {\bibinfo {author} {\bibfnamefont {J.~P.}\ \bibnamefont
  {Perdew}}, \bibinfo {author} {\bibfnamefont {K.}~\bibnamefont {Burke}}, \
  and\ \bibinfo {author} {\bibfnamefont {M.}~\bibnamefont {Ernzerhof}},\ }\href
  {\doibase 10.1103/PhysRevLett.77.3865} {\bibfield  {journal} {\bibinfo
  {journal} {Phys. Rev. Lett.}\ }\textbf {\bibinfo {volume} {77}},\ \bibinfo
  {pages} {3865} (\bibinfo {year} {1996})}\BibitemShut {NoStop}%
\bibitem [{\citenamefont {Rossi}\ and\ \citenamefont
  {Ferrando}(2009)}]{rossi2009BHMC}%
  \BibitemOpen
  \bibfield  {author} {\bibinfo {author} {\bibfnamefont {G.}~\bibnamefont
  {Rossi}}\ and\ \bibinfo {author} {\bibfnamefont {R.}~\bibnamefont
  {Ferrando}},\ }\href {\doibase 10.1088/0953-8984/21/8/084208} {\bibfield
  {journal} {\bibinfo  {journal} {J. Phys.: Condens. Matter}\ }\textbf
  {\bibinfo {volume} {21}},\ \bibinfo {pages} {084208} (\bibinfo {year}
  {2009})}\BibitemShut {NoStop}%
\bibitem [{\citenamefont {Plimpton}(1995)}]{lammps}%
  \BibitemOpen
  \bibfield  {author} {\bibinfo {author} {\bibfnamefont {S.}~\bibnamefont
  {Plimpton}},\ }\href {\doibase 10.1006/jcph.1995.1039} {\bibfield  {journal}
  {\bibinfo  {journal} {J. Comp. Phys.}\ }\textbf {\bibinfo {volume} {117}},\
  \bibinfo {pages} {1} (\bibinfo {year} {1995})}\BibitemShut {NoStop}%
\bibitem [{\citenamefont {Faken}\ and\ \citenamefont
  {Jonssonn}(1994)}]{cna1994}%
  \BibitemOpen
  \bibfield  {author} {\bibinfo {author} {\bibfnamefont {D.}~\bibnamefont
  {Faken}}\ and\ \bibinfo {author} {\bibfnamefont {H.}~\bibnamefont
  {Jonssonn}},\ }\href {\doibase 10.1016/0927-0256(94)90109-0} {\bibfield
  {journal} {\bibinfo  {journal} {Comput. Mater. Sci.}\ }\textbf {\bibinfo
  {volume} {2}},\ \bibinfo {pages} {279} (\bibinfo {year} {1994})}\BibitemShut
  {NoStop}%
\bibitem [{\citenamefont {Roncaglia}, \citenamefont {Rapetti},\ and\
  \citenamefont {Ferrando}(2021)}]{Roncaglia2021pccp}%
  \BibitemOpen
  \bibfield  {author} {\bibinfo {author} {\bibfnamefont {C.}~\bibnamefont
  {Roncaglia}}, \bibinfo {author} {\bibfnamefont {D.}~\bibnamefont {Rapetti}},
  \ and\ \bibinfo {author} {\bibfnamefont {R.}~\bibnamefont {Ferrando}},\
  }\href {\doibase 10.1039/D1CP02143E} {\bibfield  {journal} {\bibinfo
  {journal} {Phys. Chem. Chem. Phys.}\ }\textbf {\bibinfo {volume} {23}},\
  \bibinfo {pages} {23325} (\bibinfo {year} {2021})}\BibitemShut {NoStop}%
\bibitem [{\citenamefont {Settem}(2020)}]{settem2020chiralCuCore}%
  \BibitemOpen
  \bibfield  {author} {\bibinfo {author} {\bibfnamefont {M.}~\bibnamefont
  {Settem}},\ }\href {\doibase 10.1016/j.jallcom.2020.155816} {\bibfield
  {journal} {\bibinfo  {journal} {J. Alloys Compd.}\ }\textbf {\bibinfo
  {volume} {844}},\ \bibinfo {pages} {155816} (\bibinfo {year}
  {2020})}\BibitemShut {NoStop}%
\bibitem [{\citenamefont {Rossi}\ and\ \citenamefont
  {Ferrando}(2007)}]{polyDh2007}%
  \BibitemOpen
  \bibfield  {author} {\bibinfo {author} {\bibfnamefont {G.}~\bibnamefont
  {Rossi}}\ and\ \bibinfo {author} {\bibfnamefont {R.}~\bibnamefont
  {Ferrando}},\ }\href {\doibase 10.1088/0957-4484/18/22/225706} {\bibfield
  {journal} {\bibinfo  {journal} {Nanotechnology}\ }\textbf {\bibinfo {volume}
  {18}},\ \bibinfo {pages} {225706} (\bibinfo {year} {2007})}\BibitemShut
  {NoStop}%
\bibitem [{\citenamefont {Kostko}\ \emph {et~al.}(2005)\citenamefont {Kostko},
  \citenamefont {Morgner}, \citenamefont {Hoffmann},\ and\ \citenamefont {von
  Issendorf}}]{pesNaCu2005}%
  \BibitemOpen
  \bibfield  {author} {\bibinfo {author} {\bibfnamefont {O.}~\bibnamefont
  {Kostko}}, \bibinfo {author} {\bibfnamefont {N.}~\bibnamefont {Morgner}},
  \bibinfo {author} {\bibfnamefont {M.~A.}\ \bibnamefont {Hoffmann}}, \ and\
  \bibinfo {author} {\bibfnamefont {B.}~\bibnamefont {von Issendorf}},\ }\href
  {\doibase 10.1140/epjd/e2005-00099-3} {\bibfield  {journal} {\bibinfo
  {journal} {Eur. Phys. J. D}\ }\textbf {\bibinfo {volume} {34}},\ \bibinfo
  {pages} {133} (\bibinfo {year} {2005})}\BibitemShut {NoStop}%
\bibitem [{\citenamefont {Kostko}(2007)}]{olegThesis}%
  \BibitemOpen
  \bibfield  {author} {\bibinfo {author} {\bibfnamefont {O.}~\bibnamefont
  {Kostko}},\ }\emph {\bibinfo {title} {Photoelectron spectroscopy of
  mass-selected sodium, coinage metal and divalent metal cluster anions}},\
  \href@noop {} {\bibinfo {type} {{Ph.D.} thesis}},\ \bibinfo  {school}
  {Albert-Ludwigs-Universität Freiburg} (\bibinfo {year} {2007})\BibitemShut
  {NoStop}%
\bibitem [{\citenamefont {Kostko}\ \emph {et~al.}(2007)\citenamefont {Kostko},
  \citenamefont {Huber}, \citenamefont {Moseler},\ and\ \citenamefont {von
  Issendorff}}]{oleg92}%
  \BibitemOpen
  \bibfield  {author} {\bibinfo {author} {\bibfnamefont {O.}~\bibnamefont
  {Kostko}}, \bibinfo {author} {\bibfnamefont {B.}~\bibnamefont {Huber}},
  \bibinfo {author} {\bibfnamefont {M.}~\bibnamefont {Moseler}}, \ and\
  \bibinfo {author} {\bibfnamefont {B.}~\bibnamefont {von Issendorff}},\ }\href
  {\doibase 10.1103/PhysRevLett.98.043401} {\bibfield  {journal} {\bibinfo
  {journal} {Phys. Rev. Lett.}\ }\textbf {\bibinfo {volume} {98}},\ \bibinfo
  {pages} {043401} (\bibinfo {year} {2007})}\BibitemShut {NoStop}%
\bibitem [{\citenamefont {Kohn}\ and\ \citenamefont
  {Sham}(1965)}]{ldaXCFunctional}%
  \BibitemOpen
  \bibfield  {author} {\bibinfo {author} {\bibfnamefont {W.}~\bibnamefont
  {Kohn}}\ and\ \bibinfo {author} {\bibfnamefont {L.~J.}\ \bibnamefont
  {Sham}},\ }\href {\doibase 10.1103/PhysRev.140.A1133} {\bibfield  {journal}
  {\bibinfo  {journal} {Phys. Rev.}\ }\textbf {\bibinfo {volume} {140}},\
  \bibinfo {pages} {A1133} (\bibinfo {year} {1965})}\BibitemShut {NoStop}%
\bibitem [{\citenamefont {Perdew}\ \emph {et~al.}(2009)\citenamefont {Perdew},
  \citenamefont {Ruzsinszky}, \citenamefont {Csonka}, \citenamefont {Vydrov},
  \citenamefont {Scuseria}, \citenamefont {Constantin}, \citenamefont {Zhou},\
  and\ \citenamefont {Burke}}]{pbesolXCFunctional}%
  \BibitemOpen
  \bibfield  {author} {\bibinfo {author} {\bibfnamefont {J.~P.}\ \bibnamefont
  {Perdew}}, \bibinfo {author} {\bibfnamefont {A.}~\bibnamefont {Ruzsinszky}},
  \bibinfo {author} {\bibfnamefont {G.~I.}\ \bibnamefont {Csonka}}, \bibinfo
  {author} {\bibfnamefont {O.~A.}\ \bibnamefont {Vydrov}}, \bibinfo {author}
  {\bibfnamefont {G.~E.}\ \bibnamefont {Scuseria}}, \bibinfo {author}
  {\bibfnamefont {L.~A.}\ \bibnamefont {Constantin}}, \bibinfo {author}
  {\bibfnamefont {X.}~\bibnamefont {Zhou}}, \ and\ \bibinfo {author}
  {\bibfnamefont {K.}~\bibnamefont {Burke}},\ }\href {\doibase
  110.1103/PhysRevLett.100.136406} {\bibfield  {journal} {\bibinfo  {journal}
  {Phys. Rev. Lett.}\ }\textbf {\bibinfo {volume} {100}},\ \bibinfo {pages}
  {136406} (\bibinfo {year} {2009})}\BibitemShut {NoStop}%
\bibitem [{\citenamefont {Williams}, \citenamefont {Mishin},\ and\
  \citenamefont {Hamilton}(2006)}]{potCuAg}%
  \BibitemOpen
  \bibfield  {author} {\bibinfo {author} {\bibfnamefont {P.~L.}\ \bibnamefont
  {Williams}}, \bibinfo {author} {\bibfnamefont {Y.}~\bibnamefont {Mishin}}, \
  and\ \bibinfo {author} {\bibfnamefont {J.~C.}\ \bibnamefont {Hamilton}},\
  }\href {\doibase 10.1088/0965-0393/14/5/002} {\bibfield  {journal} {\bibinfo
  {journal} {Modelling Simul. Mater. Sci. Eng.}\ }\textbf {\bibinfo {volume}
  {14}},\ \bibinfo {pages} {817} (\bibinfo {year} {2006})}\BibitemShut
  {NoStop}%
\bibitem [{\citenamefont {Grochola}, \citenamefont {Russo},\ and\ \citenamefont
  {Snook}(2005)}]{potAuEAM}%
  \BibitemOpen
  \bibfield  {author} {\bibinfo {author} {\bibfnamefont {G.}~\bibnamefont
  {Grochola}}, \bibinfo {author} {\bibfnamefont {S.~P.}\ \bibnamefont {Russo}},
  \ and\ \bibinfo {author} {\bibfnamefont {I.~K.}\ \bibnamefont {Snook}},\
  }\href {\doibase 10.1063/1.2124667} {\bibfield  {journal} {\bibinfo
  {journal} {J. Chem. Phys.}\ }\textbf {\bibinfo {volume} {123}},\ \bibinfo
  {pages} {204719} (\bibinfo {year} {2005})}\BibitemShut {NoStop}%
\end{thebibliography}%
\bibliographystyle{aipnum4-1}

\clearpage

\section*{Supplementary Material}

\renewcommand{\figurename}{\textbf{Fig. S}}
\renewcommand{\tablename}{\textbf{Table S}}

\renewcommand{\thetable}{\arabic{table}}

\setcounter{figure}{0}
\setcounter{table}{0}

\begin{figure}[!ht]
\centering
  \includegraphics[width=0.5\textwidth]{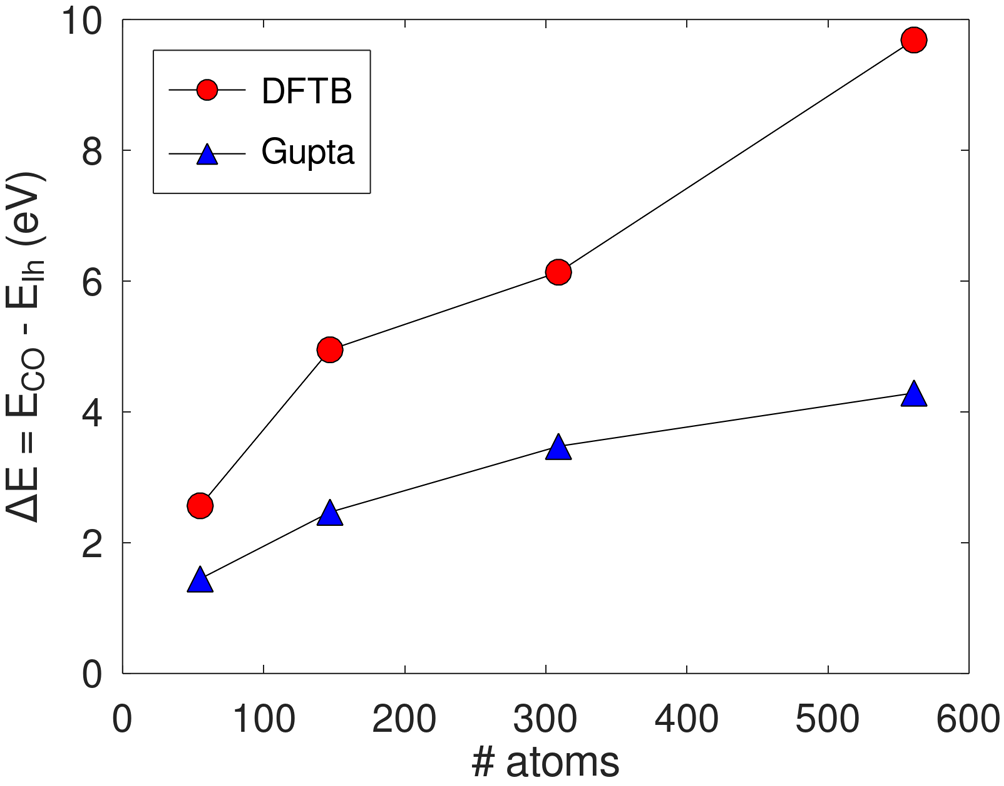}
  \caption{Comparison of $\Delta$E, energy difference between cuboctahedron (CO) and icosahedron (Ih) for Ag clusters consisting of 55, 147, 309, and, 561 atoms calculated using Gupta potential and density functional tight binding (DFTB) calculations. The values corresponding to DFTB are taken from ref. 62 of the main manuscript.}
  \label{fgr:Ag_gupta_vs_DFTB}
\end{figure}

\subsection*{Parameters of HSA and PTMD}

\begin{table}[!b]
\small
\caption{Number of local minima used for HSA analysis of Cu nanoclusters. At each size, the minima were collected up to an energy (E\textsubscript{cutoff}) above the global minimum. E\textsubscript{cutoff} values at the sizes 90, 147, and 201 are 1.2 eV, 2.5 eV, and 1.5 eV respectively.}
\centering
{\def\arraystretch{1.15}
\begin{tabular*}{0.55\textwidth}{@{\extracolsep{\fill}}l|lll}
\hline
\multirow{2}{2em}{Structure} & \multicolumn{3}{c}{Size} \\
& {{90}} & {{147}} & {{201}}\\
\hline
{{amorphous}} & {{22}} & {{$-$}} & {{$-$}} \\
{{fcc}} & {{14}} & {{1}} & {{2}} \\
{{twin}} & {{2077}} & {{42}} & {{1339}} \\
{{Ih}} & {{1825}} & {{1629}} & {{2689}} \\
{{Dh}} & {{1639}} & {{89}} & {{7174}} \\
{{mix}} & {{5543}} & {{2}} & {{$-$}} \\
\hline
{{all}} & {{11120}} & {{1763}} & {{11204}} \\
\hline
\end{tabular*}
}
\label{tab:HSAconfigsCu}
\end{table}

\begin{table}[!t]
\small
\caption{Number of local minima used for HSA analysis of Ag nanoclusters. At each size, the minima were collected up to an energy (E\textsubscript{cutoff}) above the global minimum. E\textsubscript{cutoff} values at the sizes 90, 147, and 201 are 1.0 eV, 2.5 eV, and 1.5 eV respectively.}
\centering
{\def\arraystretch{1.15}
\begin{tabular*}{0.55\textwidth}{@{\extracolsep{\fill}}l|lll}
\hline
\multirow{2}{2em}{Structure} & \multicolumn{3}{c}{Size} \\
& {{90}} & {{147}} & {{201}}\\
\hline
{{amorphous}} & {{$-$}} & {{$-$}} & {{$-$}} \\
{{fcc}} & {{117}} & {{$-$}} & {{32}} \\
{{twin}} & {{2724}} & {{98}} & {{891}} \\
{{Ih}} & {{854}} & {{2435}} & {{$-$}} \\
{{Dh}} & {{2882}} & {{444}} & {{8956}} \\
{{mix}} & {{4019}} & {{53}} & {{$-$}} \\
\hline
{{all}} & {{10596}} & {{3030}} & {{9879}} \\
\hline
\end{tabular*}
}
\label{tab:HSAconfigsAg}
\end{table}

\begin{table}[!t]
\small
\caption{Number of local minima used for HSA analysis of Au nanoclusters. At each size, the minima were collected up to an energy (E\textsubscript{cutoff}) above the global minimum. E\textsubscript{cutoff} values at the sizes 90, 147, and 201 are 1.0 eV, 1.5 eV, and 1.0 eV respectively. These data are taken from \cite{settem2022AuPTMD}.}
\centering
{\def\arraystretch{1.15}
\begin{tabular*}{0.55\textwidth}{@{\extracolsep{\fill}}l|lll}
\hline
\multirow{2}{2em}{Structure} & \multicolumn{3}{c}{Size} \\
& {{90}} & {{147}} & {{201}}\\
\hline
{{amorphous}} & {{69}} & {{$-$}} & {{$-$}} \\
{{fcc}} & {{316}} & {{1382}} & {{728}} \\
{{twin}} & {{3107}} & {{3651}} & {{1316}} \\
{{Ih}} & {{1}} & {{2444}} & {{$-$}} \\
{{Dh}} & {{994}} & {{8554}} & {{8583}} \\
{{mix}} & {{6632}} & {{1919}} & {{$-$}} \\
\hline
{{all}} & {{11119}} & {{17950}} & {{10627}} \\
\hline
\end{tabular*}
}
\label{tab:HSAconfigsAu}
\end{table}

For the HSA analysis, local minima were selected by applying an energy cutoff (E\textsubscript{cutoff}) which was adjusted to have roughly 10000 configurations. This resulted in E\textsubscript{cutoff} values in the range 1.0 eV to 1.5 eV. However, in the case of Cu$_{147}$ and Ag$_{147}$, Ih spans the entire temperature range. Even a high E\textsubscript{cutoff} of 2.5 eV results in $\sim$ 1700, $\sim$ 3000 local minima for Cu$_{147}$, Ag$_{147}$ respectively. The number of local minima used per each motif and the energy cutoffs of Cu, Ag, and Au are provided in the Tables S1, S2, and S3 respectively. PTMD parameters (\# of replicas and temperature of replicas) are provided in the Table S4.

\begin{table*}[!t]
\small
\caption{\# of replicas and replica temperatures used for PTMD simulations of Cu, Ag, and Au nanoclusters.}
\centering
{\def\arraystretch{1.15}
\begin{tabular*}{15cm}{@{\extracolsep{\fill}}p{2.7cm}|p{1.7cm}|p{10.6cm}}
\hline
{Metal cluster} & {\# replicas} & {Replica temperatures (K)}\\
\hline
Cu$_{90}$ & {37} & {300, 315, 331, 347, 365, 383, 402, 423, 444, 466, 490, 514, 540, 550, 558, 567, 575, 584, 592, 601, 609, 617, 626, 634, 643, 651, 659, 668, 676, 685, 693, 702, 710, 720, 746, 772, 800}\\
\hline
Cu$_{147}$ & {40} & {300,   310,   321,   332,   343,   355,   367,   379,   392,   405,   419,   433,   448,   463,   479,   495,   512,   530,   548,   566,   586,   605,   626,   647,   669,   692,   716,   740,   750,   758   767,   775,   783,   792,   800,   810,   832,   854,   877,   900}\\
\hline
Cu$_{201}$ & {40} & {300,   311,   322,   333,   345,   357,   369,   382,   396,   410,   424,   439,   455,   471,   488,   505,   523,   541,   560,   580,   601,   622,   644,   666,   690,   700,   709,   718,   727,   736,   744,   753,   762,   771,   780,   790,   816,   843,   871,   900}\\
\hline
Ag$_{90}$ & {30} & {250,   264,   280,   296,   313,   331,   350,   370,   392,   414,   438,   463,   471, 479, 486, 493,   500,   507,   513,   520,   527,   533,   540,   547,   553,   560,   570,   596,   622,   650}\\
\hline
Ag$_{147}$ & {32} & {300,   313,   326,   339,   354,   369,   384,   400,   417,   435,   453,   472,   492,   513,   534,   557,   580,   605,   630,   640,   647,   654,   661,   669,   676,   683,   690,   700,   724,   748,   774,   800}\\
\hline
Ag$_{201}$ & {36} & {300,   312,   324,   337,   351,   365,   379,   394,   410,   426,   443,   461,   479,   498,   518,   539,   560,   583,   606,   630,   640,   646,   652,   658,   664,   670,   676,   682,   688,   694,   700,   710,   732,   754,   776,   800}\\
\hline
Au$_{90}$ & {36} & {250, 263, 275, 288, 300, 309, 319, 329, 339, 350, 357, 363, 370, 376, 383, 389, 396, 402, 409, 415, 422, 428, 435, 441, 448, 454, 461, 467, 474, 480, 487, 493, 500, 517, 533, 550}\\
\hline
Au$_{147}$ & {24} & {300, 314, 329, 345, 361, 378, 396, 415, 434, 455, 476, 482, 488, 493, 499, 505, 511, 516, 522, 528, 546, 564, 582, 600}\\
\hline
Au$_{201}$ & {32} & {300, 312, 324, 336, 349, 363, 377, 391, 406, 422, 438, 455, 473, 491, 510, 516, 521, 527, 533, 539, 544, 550, 556, 561, 567, 573, 579, 584, 590, 610, 630, 650}\\
\hline
\end{tabular*}
}
\label{tab:ptmd_params}
\end{table*}

\clearpage

\subsection*{Structural distribution of Au nanoclusters}
In Fig. S2, we replot the structural distribution of Au$_{90}$, Au$_{147}$, and Au$_{201}$ clusters using the data reported previously \cite{settem2022AuPTMD} for the purpose of comparison with Cu and Ag clusters.

\begin{figure}[!h]
\centering
  \includegraphics[width=1.0\textwidth]{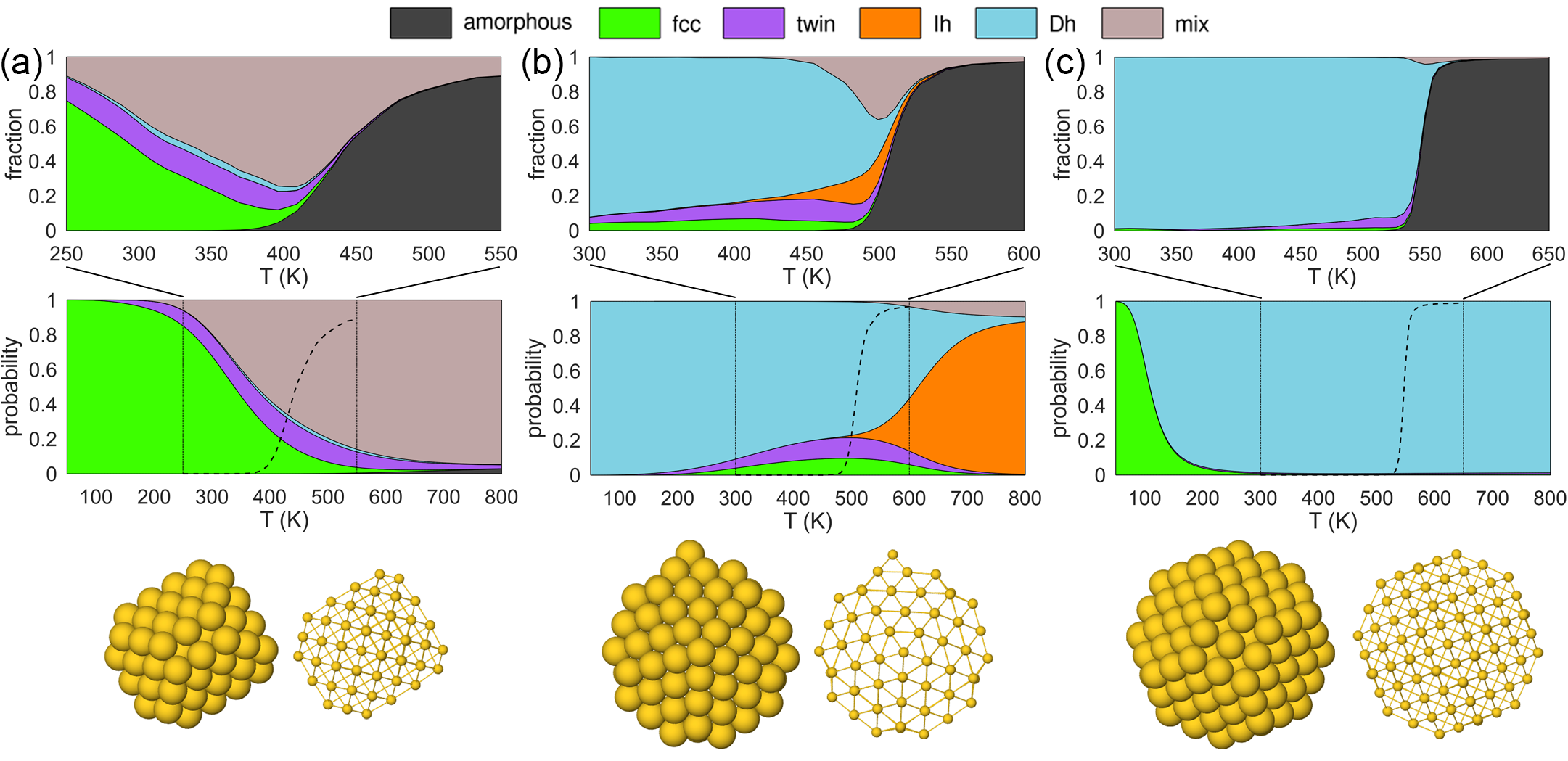}
  \caption{Structural distribution of (a) Au$_{90}$, (b) Au$_{147}$, and (c) Au$_{201}$. PTMD, HSA results are shown in the top and middle rows. Global minimum structures are shown in the bottom row. In the HSA results, for comparison, we report with vertical lines  the range of PTMD temperatures and with a dashed line the fraction of amorphous structures calculated from PTMD simulations.}
  \label{fgr:au_ptmd_hsa}
\end{figure}



\end{document}